\documentclass[letterpaper,11pt]{article}

\usepackage{amsmath,amssymb}
\usepackage{array}
\usepackage{float}
\usepackage{amsmath,amsfonts,graphicx,bm}
\usepackage{amsmath,amsfonts,graphicx,bbm,multicol}

\newcommand{\mP}{{\bar M}_{P}}

\begin{document}

\baselineskip=18pt

\newcommand{\eps}{\epsilon}
\newcommand{\pslash}{\!\not\! p}
\newcommand{\I}{\rm 1\kern-.24em l} 
\newcommand{\Tr}{\mathop{\rm Tr}}
\def\mz{M_Z}
\def\mw{M_W}
\def\mh{M_h}
\def\ap{A_1}
\def\zp{Z_1}
\def\zx{{Z_X}_1}
\def\zpri{{Z^\prime}}
\def\zpt{\tilde{Z}_1}
\def\zxt{\tilde{Z}_{X1}}
\def\map{M_{A_1}}
\def\mzp{M_{Z_1}}
\def\mzx{M_{{Z_X}_1}}
\def\mzpri{M_{Z^\prime}}
\def\wpL{W_{L_1}}
\def\wpR{W_{R_1}}
\def\wpri{{W^\prime}}
\def\wpLt{\tilde{W}_{L_1}}
\def\wpRt{\tilde{W}_{R_1}}
\def\wpLnt{W'_L}
\def\wpRnt{W'_R}
\def\mwpL{M_{W_{L_1}}}
\def\mwpR{M_{W_{R_1}}}
\def\mwpri{M_{W^\prime}}

\def\beq{\begin{equation}}
\def\eeq{\end{equation}}
\def\bea{\begin{eqnarray}}
\def\eea{\end{eqnarray}}
\def\bmat{\begin{pmatrix}}
\def\emat{\end{pmatrix}}
\def\to{\rightarrow}
\def\check{{\bf Check: }}
\def\todo{{\bf To do: }}
\def\gev{\rm GeV}
\def\tev{\rm TeV}
\def\fbi{\rm fb^{-1}}
\def\lsim{\mathrel{\raise.3ex\hbox{$<$\kern-.75em\lower1ex\hbox{$\sim$}}}}
\def\gsim{\mathrel{\raise.3ex\hbox{$>$\kern-.75em\lower1ex\hbox{$\sim$}}}}
%
\newcommand{ \slashchar }[1]{\setbox0=\hbox{$#1$}   
   \dimen0=\wd0                                     
   \setbox1=\hbox{/} \dimen1=\wd1                   
   \ifdim\dimen0>\dimen1                            
      \rlap{\hbox to \dimen0{\hfil/\hfil}}          
      #1                                            
   \else                                            
      \rlap{\hbox to \dimen1{\hfil$#1$\hfil}}       
      /                                             
   \fi}                                             %
\def\ptmiss{\slashchar{p}_{T}}
\def\etmiss{\slashchar{E}_{T}}
\providecommand{\tabularnewline}{\\}


\thispagestyle{empty}
\vspace{20pt}
\font\cmss=cmss10 \font\cmsss=cmss10 at 7pt

\begin{flushright}
UMD-PP-08-014 \\
MADPH-08-1520 \\
UCD-2008-03
\end{flushright}


\begin{center}
{\Large \textbf
{LHC Signals for Warped Electroweak Charged Gauge Bosons}}
\end{center}

\vspace{5pt}

\begin{center}
{\large 
Kaustubh Agashe$\, ^{a}$, 
Shrihari Gopalakrishna$\, ^{b}$,
Tao Han$\, ^{c}$, 
Gui-Yu Huang$\, ^{c,d}$, 
Amarjit Soni$\, ^{b}$
} \vspace{20pt}

$^{a}$\textit{Maryland Center for Fundamental Physics,
     Department of Physics,
     University of Maryland,
     College Park, MD 20742, USA.}
\\
$^{b}$\textit{Brookhaven National Laboratory,
Upton, NY 11973, USA.}
\\
$^{c}$\textit{Department of Physics, University of
Wisconsin, Madison, WI 53706, USA.}
\\
$^{d}$\textit{Department of Physics, University of California at Davis, Davis, CA 95616, USA.}
\end{center}

\begin{center}
\textbf{Abstract}
\end{center}

We study signals at the Large Hadron Collider
(LHC) for the Kaluza-Klein (KK) excitations of electroweak {\em charged} gauge bosons in 
the framework of the Standard Model (SM) fields propagating in the bulk of a 
warped extra dimension.
Such a scenario can solve both the Planck-weak and flavor
hierarchy problems of the SM. There are two such charged states
in this scenario with couplings to light quarks and leptons being 
suppressed relative to those in the SM, whereas the couplings to top/bottom quarks are enhanced,
similar to the case of electroweak neutral gauge bosons previously studied. However, 
unlike the case of electroweak neutral gauge bosons, there is no irreducible QCD background 
(including pollution from possibly degenerate KK gluons) for decays to top $+$ bottom final state
so that this channel is useful for the discovery of the charged states.
Moreover, decays of electroweak charged gauge bosons to longitudinal $W$, $Z$ and Higgs 
are enhanced just as for the neutral bosons. However, unlike for the neutral gauge bosons, 
the purely leptonic (and hence clean) decay mode of the $WZ$ are fully reconstructible
so that the ratio of the signal to the SM (electroweak) background can potentially be
enhanced by restricting to the resonance region more efficiently. 
We show that such final states can give sensitivity to $2~(3)$ TeV masses
with an integrated luminosity of $100~(300)$ fb$^{-1}$.
We emphasize that improvements in discriminating a QCD-jet from a highly boosted hadronically
decaying $W$, and a highly boosted top-jet from a bottom-jet 
will enhance the reach for these KK particles, and that the signals we study 
for the warped extra dimensional model might actually be applicable also to a wider class of 
non-supersymmetric models of electroweak symmetry breaking.

\vspace{5pt} {\small \noindent

}

\vfill\eject
\noindent


\section{Introduction}

The era of the Large Hadron Collider (LHC) is 
upon us! Experiments at the LHC are highly expected  to 
shed light on the mechanism of electroweak symmetry breaking (EWSB).
In particular, various extensions of the
Standard Model (SM) have been proposed to solve
the problem in the SM of the hierarchy between the
Planck and electroweak scales. Such models predict
the existence of new particles at the
weak (or TeV) scale which are likely to be accessible to the LHC.

In the present work, 
we focus on one such extension of the
SM, the Randall-Sundrum model (RS1) \cite{rs1}
with all the SM fields propagating
in the bulk of a warped extra dimension \cite{bulk, gn, gp}. Such a framework
can address the
flavor hierarchy problem of the SM as well. 
The versions of this framework
with a grand unified gauge symmetry in the bulk 
can naturally lead to precision unification of the three
SM gauge couplings \cite{Agashe:2005vg} and 
%
%
a candidate for the 
dark matter of the universe 
(the latter from
requiring longevity of the proton)
\cite{Agashe:2004ci}. 
The
new particles in this framework are Kaluza-Klein (KK) 
excitations of all SM fields with masses at $\sim {\rm TeV}$ scale.
So far, 
%
%
studies of the LHC
signals from direct production of the radion
\cite{radion}, KK gluon \cite{kkgluon, Lillie:2007yh,
otherkkgluon, Carena:2007tn}, 
graviton \cite{Fitzpatrick:2007qr}, {\em neutral} electroweak gauge bosons 
\cite{Agashe:2007ki}, (heavy) fermions \cite{Davoudiasl:2007wf} and finally
{\em light} KK fermions present in some models
with extended $5D$ gauge symmetries \cite{Dennis:2007tv} in such a 
framework have been performed 
(see Ref.~\cite{Davoudiasl:2007wf}
for an overview and Refs.~\cite{otherLHC} for related studies in other
set-ups within the warped extra dimensional framework).

However,
there are some challenging 
aspects of this collider phenomenology as follows.
Firstly,  
the KK mass scale is constrained to be at least a  
few TeV by the electroweak and flavor precision tests
in part due to the absence of a parity symmetry
(analogous to $R$-parity in SUSY), allowing tree-level exchanges
to contribute to the precision observables.
In addition, the 
constituents of the proton (or SM gauge bosons
and light fermions in general) couple weakly to the
KK states, whereas the KK states mostly decay to top quarks and 
longitudinal $W/Z$/Higgs
due to a larger coupling to these states. 
As a result, the golden decay channels such as resonant signals of dileptons or diphotons
are suppressed. Finally, given the few TeV KK mass, the
top quarks/$W$/$Z$ resulting from the decays of these KK states
are highly boosted, creating problems in their identification
due to collimation of their decay products.

In light of this situation,  it is
necessary to study as many LHC probes of this framework
as possible, especially since there might not be
a single ``smoking gun'' for this framework, i.e., 
a variety of channels can complement
each other as far as detecting this framework
at the LHC is concerned. 
In particular, the most widely studied particle is the 
KK gluon which decays only to jetty final states, but has the 
largest cross-section due to the QCD coupling
(assuming the same mass for all KK particles as in the simplest models). 
It was found that the LHC reach can be $\sim 4$ TeV, using techniques
designed specifically to identify highly boosted top quarks.
However, it is good to
have channels with no jets if possible, since in general 
such modes are cleaner in the LHC experimental environment.

Also, it is obviously important to explore the feasibility of searching for the
electroweak (EW) KK states (i.e., excitations of $\gamma$,  $W$ and $Z$)
at the LHC. In fact, decays of EW KK states to (longitudinal) $W/Z$ and Higgs offer
a possibility for clean final states if these
SM particles decay to leptons
(note that the direct decays to leptons/photons are suppressed
as mentioned above).
However, the decay of the KK
$Z$ to $WW$ followed by leptonic decays of both $W$'s has
two neutrinos in the final state
so that the invariant mass of $W$ pair cannot be effectively 
reconstructed, making it harder to identify the signal and to reduce 
the continuum SM $WW$ background. If one $W$ decays instead to a pair of quarks 
(we call it ``semileptonic'' decay of the $WW$ gauge boson pair), then
the problem is that the two jets from the $W$ are collimated, introducing a larger 
QCD background from $W +$ jet. 
A similar analysis applies to decays of the KK $Z$ to the $Zh$
final state.
Finally, one could utilize decays to top pairs for detecting the
KK $Z$ using techniques to identify
boosted tops developed for detecting the KK gluon, but 
this channel is swamped by decays of the KK gluon to (i.e., resonant production of) top pairs, 
if not by the SM $t \bar{t}$ continuum background.  

Given this situation, the KK $W$ can provide (a priori) a couple of advantages:
\begin{itemize}
\item
The decays to $WZ$ followed by (clean) leptonic decays
of both $W$ and $Z$ can be more effectively reconstructed due to the presence of only one
neutrino. 
\end{itemize}

The semileptonic decays of $WZ$ or $Wh$ face similar challenges
to those of KK $Z$, namely, QCD $Z/W +$ jet 
background and we can use a jet mass cut
in order to reduce this background, i.e., to distinguish
a $W/Z$-jet from a QCD jet.

%
\begin{itemize}
\item
The decays of KK $W$ to $t \bar{b}$ do not have the contamination 
from the KK gluon as in the case of KK $Z$.
\end{itemize}
With this background, we are thus motivated to study signals
for the KK excitation of the SM $W$ in this paper. With detailed parton-level 
simulations for the signal and SM backgrounds,
we find the reach for this particle to be $2 \; (3)$ TeV with $\sim 100
\; (300)$ fb$^{-1}$ luminosity in the $t \bar{b}$, $WZ$ and $Wh$ channels, 
a discovery potential which is 
roughly similar to that for the KK $Z$, found earlier in 
\cite{Agashe:2007ki}. 
The reason for the similar (although slightly better) reach for the KK $W$ as for the KK $Z$ in spite
of the expected above two advantages for the former are that, firstly,
the BR to leptons for the SM $Z$ is smaller than that for the $W$, making the final 
significance of the leptonic decays of the $WZ$ from the
KK $W$ not much better than in the case of the
purely leptonic decays of the $W$ pairs from the KK $Z$.
Secondly, we find that a highly boosted top quark can fake a bottom quark
so that QCD or KK gluon $t \bar{t}$ pairs do manifest as 
{\em reducible} backgrounds to $t \bar{b}$ signal from the KK $W$. 
Once again, we use a jet mass cut, this time
to discriminate between a $t$-jet and a $b$-jet.
Anticipating
more dedicated analyses in regard to vetoing $t \bar{t}$
background from KK gluon decays,
we believe  that it is possible to make the reach in KK $W$ {\em better}
than KK $Z$ via the $t \bar{b}$ channel, perhaps comparable to KK gluon.
Similarly, further improvements in the jet mass technique
of distinguishing a $W/Z$-jet from a QCD jet
or the development of new ones to reduce this
QCD background can increase the reach for
{\em both} KK $W$ and $Z$ in semileptonic $WW/WZ$ 
decays of these KK modes.

The outline of the paper
is as follows. In Sec.~\ref{review}, we begin with a brief review
of the theory of a warped extra dimension and the LHC signals
for the KK states, including an outline of the various cases we consider
for the study of 
the electroweak charged gauge bosons.
We present the 
total decay widths of $W^{ \prime }$'s
and the branching ratios to various channels 
in Sec.~\ref{decay}.
In Sec.~\ref{signal}, we calculate the
production cross-sections of the charged gauge bosons at the LHC
and present a detailed
analysis of how to obtain signals for these states. The framework of
warped extra
dimension is conjectured to be dual to four-dimensional ($4D$) 
strong dynamics triggering electroweak symmetry breaking,
as in technicolor or composite Higgs models. 
In Sec.~\ref{TC}, we then compare the signals that we studied
in a warped extra dimension to the signals for technicolor models
discussed previously (since 1990's). Further discussions and 
conclusions are presented in Sec.~\ref{conclude}, where
we argue that many of the signals that we study here (including
the electroweak neutral gauge boson case
studied earlier) might be
applicable to a wider class of non-supersymmetric models
of EWSB.
Finally,
two Appendices are included at the end to provide further details
for the model, including the couplings of the $W^{\prime}$ states.

\section{Review of Warped Extra Dimension}
\label{review}

The framework consists of a slice of anti-de Sitter
space in five dimensions (AdS$_5$), where (due to the
warped geometry) the effective $4D$ mass scale is dependent
on position in the extra dimension.
The $4D$
graviton, i.e., the zero-mode
of the $5D$ graviton, is automatically localized
at one end of the extra dimension (called the Planck/UV brane).
If the Higgs sector is localized at the other end (in fact
with SM Higgs originating as 5th
component of a $5D$ gauge field
($A_5$) it is automatically so \cite{Contino:2003ve}), then the warped geometry
naturally generates the Planck-weak hierarchy.
Specifically, TeV $\sim \mP
e^{ - k \pi r_c }$, where $\mP$ is
the reduced $4D$ Planck scale,
$k$ is the AdS$_5$ curvature scale and $r_c$ is the proper
size of the extra dimension. The crucial 
point is that the required
modest size of the radius (in units of the curvature radius), i.e.,
$k r_c \sim 1 / \pi \log \left( \mP / \hbox{TeV}
\right) \sim 10$ can be 
%
%
stabilized with only a corresponding
modest tuning in the fundamental or $5D$ parameters of
the theory \cite{Goldberger:1999uk}.
Remarkably, the correspondence between
AdS$_5$  
and $4D$ conformal field theories (CFT) \cite{Maldacena:1997re}
suggests that 
the scenario with warped extra dimension is 
dual to the idea of a composite Higgs in $4D$ 
\cite{Arkani-Hamed:2000ds, Contino:2003ve}.

It was realized that with 
SM fermions propagating in the bulk, we can also
account for the hierarchy between quark 
and lepton masses and mixing angles (flavor hierarchy)
\cite{gn, gp}. The
basic idea is that light SM fermions --
which are the zero-modes of $5D$ fermions --
can be localized 
near the Planck brane, resulting in a 
small overlap with the TeV-brane localized SM Higgs, while
the top quark is localized near the TeV brane with a large 
overlap with the Higgs.
Again, the crucial point is that such vastly different profiles 
can be realized with small variations in the 
$5D$ mass parameters of fermions, i.e., without
any large hierarchies in the parameters of
the $5D$ theory.
Due to the different profiles
of the SM fermions in the extra dimension, 
flavor changing neutral
currents (FCNC) are generated
by their non-universal couplings to gauge KK 
states.
However, 
these contributions to
the FCNC's are suppressed due to an analog of the Glashow-Iliopoulos-Maiani
(GIM) mechanism of the SM, i.e. RS-GIM, 
%
%
which is ``built-in''~\cite{gp, hs, aps}.
The point is that {\em all} KK modes
(whether gauge, graviton or fermion) are localized near the
TeV or IR brane (just like the Higgs) so that non-universalities
in their couplings to SM fermions are of
same size as couplings to the Higgs.

In spite of this RS-GIM
suppression, it was shown recently \cite{Csaki:2008zd}
(see also \cite{Fitzpatrick:2007sa,
Davidson:2007si}),
that 
the constraint on the KK mass scale from
contributions of KK gluon to $\epsilon_K$ 
is quite stringent.
In particular, for the model with the
SM Higgs (strictly) localized on the TeV brane,
the limit on the KK
mass scale from $\epsilon_K$ is $\sim 10-40$ TeV,
depending on the size of the $5D$ QCD gauge coupling.
However, the phenomenology of the TeV-scale KK modes and the SM
Higgs is quite sensitive to the structure near the TeV brane
(where these particles are localized). For example, 
the SM Higgs can be the lightest mode of
a $5D$ scalar (instead of being a strictly TeV brane-localized
field), but with a
profile which is still peaked near the TeV brane
(such that the
Planck-weak hierarchy is still addressed)
i.e. a ``bulk Higgs''~\cite{Davoudiasl:2005uu}.
Moreover, the warped
geometry might deviate from pure AdS near the TeV brane which in
fact could be replaced with a ``soft wall'' \cite{soft}. Similarly,
in general, there are non-zero TeV brane-localized kinetic terms for
the bulk fields \cite{Carena:2002dz}.
Such variations of the minimal models are not likely to modify 
the constraint on KK mass scale from various precision tests 
by much more than $O(1)$ factors. However, even such
modest changes can dramatically impact the LHC signals, especially
the production cross-sections for the KK modes.

With the above motivation, the 
``two-site model'' \cite{Contino:2006nn} was proposed
as an economical
%
%
description of this framework in order to capture the
robust aspects
of the phenomenology by
effectively restricting to the SM
fields and their first KK excitations. In reference \cite{Agashe:2008uz},
it was shown that a mass scale for the new particles
as low as $\sim O(5)$ TeV is consistent with
the combination of constraints from $\epsilon_K$
and BR $\left( b \rightarrow s \gamma \right)$, and it
was suggested that 
models with a bulk Higgs can allow a similar KK scale.
In addition, mechanisms exist  
to ameliorate such constraints in a {\em parametric} manner,
for example through flavor symmetries~\cite{Cacciapaglia:2007fw, Fitzpatrick:2007sa} or
by lowering the UV-IR hierarchy \cite{Davoudiasl:2008hx}, 
as opposed to simply relying on the $O(1)$ effects mentioned
above.

Most of the studies of the KK gluon, graviton and $Z$
(and similarly our study of the KK $W$ here) focus on flavor-{\em preserving}
fermionic decays (i.e., $t \bar{t}$ for neutral and $t \bar{b}$
for charged case), except for Ref.~\cite{Aquino:2006vp}
which considers flavor-{\em violating}
decays of the KK gluon. So, it is important
to point out that the results of these studies
apply to the warped extra dimensional framework independent
of the
specific mechanism used to suppress flavor violation (beyond
that from the RS-GIM mechanism) since
the profiles and hence the (flavor -preserving) couplings remain (roughly)
the same in all these different models for suppressing flavor
violation (except in Ref.~\cite{Davoudiasl:2008hx} with UV-IR 
hierarchy being smaller than Planck-weak).
For other studies of flavor physics, see Refs.~\cite{others1, others2}.

Finally, various custodial symmetries \cite{custodial1, custodial2}
can be incorporated such that the constraints from the various 
(flavor-preserving) electroweak precision tests (EWPT)
can be satisfied for a few TeV KK scale \cite{custodial1,EWPTmodel}.
The 
bottom line is that a 
few TeV mass scale for the KK gauge bosons can be consistent with both
electroweak and flavor precision tests.

\subsection{Couplings}

Clearly, the light fermions have a small
couplings to all KK's (including graviton)
based simply on the overlaps
of the corresponding profiles, 
while the top quark and Higgs have a large coupling to the KK's.
Schematically,
neglecting effects related to EWSB, we find the following ratio of
RS1 to SM
gauge couplings:
\begin{eqnarray}
{g_{\rm RS}^{q\bar q,l\bar l\, A^{ (1) }}\over g_{\rm SM}}
&\simeq&
- \xi^{-1}\approx - {1\over5} \nonumber \\
{g_{\rm RS}^{Q^3\bar Q^3 A^{ (1) }}\over g_{\rm
    SM}},
{g_{\rm RS}^{t_R\bar t_R A^{ (1) }}\over g_{\rm
    SM}} 
& \simeq & 
1 \; \hbox{to} \; \xi \; ( \approx 5 ) \nonumber \\
{g_{\rm RS}^{ HH A^{(1)}}\over g_{\rm
    SM}}  
& \simeq & 
\xi \approx 5 \; \; \; \left( H = h, W_L, Z_L \right)
\nonumber \\
{g_{\rm RS}^{ A^{ (0) }A^{ (0) } A^{ (1) }}\over g_{\rm
    SM}}  
& \sim & 0
\label{RScouplings}
\end{eqnarray}
Here $q=u,d,s,c,b_R$, $l =$ all leptons, $Q^3= (t, b)_L$, 
and $A^{ (0) }$ ($A^{ (1) }$) correspond
to zero (first KK) states of the gauge fields. Also, 
$g_{\rm RS}^{xyz}, g_{\rm SM}$ stands for the RS1 and the three SM (i.e.,
$4D$) gauge couplings respectively.
Note that 
$H$ includes both the physical Higgs ($h$) and 
{\em un}physical Higgs, i.e., {\em longitudinal}
$W/Z$ by the equivalence theorem
(the derivative involved in this coupling is 
similar for RS1 and SM cases and hence is not shown for
simplicity). Finally, the parameter $\xi$ is
related to the Planck-weak hierarchy: $\xi \equiv \sqrt{ k \pi r_c }$.
EWSB induces mixing between EW KK states
which we discuss in App.~\ref{Coupl.APP}.

For completeness, we present 
the couplings of the KK
graviton to the SM particles. 
These couplings involve derivatives
(for the case of {\em all} SM particles),
but (apart from a factor from the overlap
of the profiles) it turns out that 
this energy-momentum dependence is
compensated (or made dimensionless) by the $\mP e^{ - k \pi r_c }\sim$ 
TeV scale, instead of 
the $\mP$-suppressed coupling to the SM graviton. Again, schematically:
\begin{eqnarray}
g_{ \rm RS }^{ q\bar q,l\bar l\, G^{ (1) } } & \sim & 
\frac{E}{ \mP e^{ - k \pi r_c } } \times 4D \; \hbox{Yukawa}
\nonumber \\
g_{ \rm RS }^{ A^{ (0) }A^{ (0) } G^{ (1) } } &
\sim & \frac{1}{ k \pi r_c }  \frac{E^2}{ \mP e^{ - k \pi r_c } }
\nonumber \\
g_{ \rm RS }^{ Q^3\bar Q^3 A^{ (1) } }, g_{ \rm RS }^{ t_R\bar t_R G^{ (1) } }
& \sim & \left( \frac{1}{ k \pi r_c } 
\; \hbox{to} \; 1 \right) 
\frac{E}{ \mP e^{ - k \pi r_c } }\nonumber \\
g_{ \rm RS }^{ H H G^{ (1) } } & \sim  & 
\frac{E^2}{ \mP e^{ - k \pi r_c } }
\end{eqnarray}
Here, $G^{ (1) }$ is the KK graviton
and the 
tensor
structure of the couplings is not shown
for simplicity.

Next, we briefly review studies of LHC signals in this
scenario.

\subsection{LHC signals}

Based on these couplings, 
and the fact that precision electroweak and flavor constraints require the
mass to be bigger than a few TeV, 
we are faced with the following challenges 
in obtaining the EW KK gauge boson signals at the LHC
from direct production of the KK modes, namely,
\begin{itemize}
\item[(i)]
Cross-section for production of these
states is suppressed to begin with
due to a small coupling to
the protons' constituents, and due to the large mass; 
\item[(ii)]
Decays to ``golden'' channels (leptons, photons)
are suppressed. Instead, the decays are 
dominated by top quark and Higgs
(including longitudinal $W/Z$); 
\end{itemize}
Also, these resonances tend to be quite
broad due to the enhanced couplings to top/Higgs.

In particular, the KK 
graviton, gluon and neutral electroweak gauge bosons all have
sizable branching ratio (BR) to decay to top {\em pairs}. Moreover,
due to the large mass (few TeV) of the KK particle,
the top quarks produced in their decays are highly boosted, 
resulting in a high degree of collimation
of the top quark's decay products. Hence it 
is a challenge to identify these top quarks. Nonetheless,
using the techniques suggested in Refs.~\cite{kkgluon, Lillie:2007yh}
(see also Refs.~\cite{otherboostedtop} for related studies and
\cite{Thaler:2008ju, Kaplan:2008ie} for recent 
%
%
developments of the
techniques for detecting highly boosted top quarks), discovery
for KK {\em gluon} up to $\sim 4$ TeV mass might be possible. However, in the
case of approximately degenerate gauge KK modes, it is still
difficult to extract the signal from top pairs for {\em electroweak} neutral
KK modes. The reason is that the top pair signal
from these states is swamped by the decays of the KK gluon 
which has a (much) larger cross-section than that of the KK electroweak
neutral gauge boson 
due to the QCD coupling and color factors, even though the
SM $t \bar{t}$ background might be smaller than the electroweak
neutral KK signal.

As mentioned above, couplings of KK's to
longitudinal $W$/$Z$ are also enhanced similarly to
top quarks (of course, only for KK graviton and electroweak
-- both neutral and charged -- KK modes) so that
decays to these modes also have a significant BR.
Such final states are 
%
%
a priori cleaner than top quarks, in
particular, since there is 
no ``pollution'' from QCD or KK gluon and there are decay channels
with no jets
in these cases.
Hence such final states might be the discovery modes
for
electroweak (both charged and neutral) and graviton KK states. However,  
we still face some challenges in discovering the
{\em neutral} electroweak (cf. charged
case below) and graviton KK states 
even with the $W/Z$ final states as follows. The purely leptonic
decay in the $WW$ channel has a small BR and moreover, the
$WW$ invariant mass cannot be fully
reconstructed due to the presence of
missing momentum
%
%
from {\em two} $\nu$'s.
Since it is difficult then to apply the mass window cut
efficiently (i.e., without reducing signal) in order to isolate
the
events in the resonance region only, the SM background tends to be
larger. It is true that the $ZZ \rightarrow
4 l$ final state can be
fully reconstructed,
but it
has an even smaller BR to leptons than the $WW$ channel and 
is available only for the the graviton (it is absent for the neutral electroweak gauge KK).

On the other hand, the semileptonic decay of the
$WW$ from the KK $Z$ and graviton has a larger BR. However,
just as for the top quarks mentioned above, the hadronic
decays of the 
highly boosted $W/Z$ pose a challenge for detection:
the  
2 jets from $W/Z$ tend to merge so that the QCD $W/Z +$ jet background
(where a QCD jet fakes 
a hadronically decaying $W/Z$) becomes significant.
Of course, this background is reducible
so that with suitable discriminators between
QCD and $W/Z$ jets such as jet mass, this channel can still be
useful \cite{Agashe:2007ki}. A similar argument applies to 
decays of KK $W$ to $WZ$.


In this paper, we study LHC signals from direct
production 
of {\em charged} electroweak KK gauge bosons
in the framework of a warped extra dimension. 
Apart from completing the study of spin-1 KK's, our
motivation for this study is that these states
possess some new features relative
to KK {\em neutral} electroweak gauge boson and graviton, namely, 
\begin{itemize}
\item[(i)]
The fully leptonic (and hence clean) decay mode of the $WZ$ channel
can be fully reconstructed\footnote{assuming
that the missing momentum
%
%
from the neutrino combined with the lepton 
forms a $W$, or assuming that the neutrino 3-momentum is collinear
with that of the lepton due to the large boost of the $W$
in the lab frame.} due to the presence of only {\em one} $\nu$.
Hence, it is expected that the signal to background ratio can be
enhanced efficiently 
%
%
by a suitable cut on the
$WZ$ invariant mass, namely, by simply requiring this mass
to lie in the resonance region (unlike
for the neutral case
discussed above).
Moreover, this final state for the $WZ$ has
a larger BR than the $ZZ \rightarrow 4 l$ case
for the KK graviton (although leptonic BR of $WZ$ is smaller than
that of the fully leptonic decay of the $WW$ final state
for the KK graviton and {\em neutral} electroweak
gauge boson).

The issues with the semileptonic decay of $WZ$ will be
similar to that in the neutral electroweak gauge boson case.

\item[(ii)]
Decays to top $+$ bottom final state of electroweak
charged gauge bosons  can be also reconstructed even for the leptonic
decay mode of the top quark.\footnote{imposing 
the on-shell conditions for $M_W$ and $m_t$.}
The irreducible SM background from the electroweak process
(single top production) can be shown to be smaller than the 
signal inside the resonance region.
Compared to the case of the neutral electroweak KK's
where the decays to top pairs have an irreducible SM background from QCD
processes, the electroweak background 
for the charged case is smaller
(and the signal cross-section for the neutral case is roughly similar
to the charged case). Moreover,  if 
the KK gauge bosons are degenerate, then an even larger background
from the KK gluon decays to top pairs
completely swamps the signal from the electroweak neutral
KK boson.

However, even for the charged case, QCD top pairs
(from KK gluon or the SM)
can be a  significant background if one top quark 
fakes a bottom quark due to collimation
of its decay products. Of course, 
techniques to distinguish a highly boosted
top from a bottom can suppress
this background, i.e., it is a {\em reducible} one.

\end{itemize}

\subsection{Overview of the charged electroweak gauge boson sector}
\label{outline}

We present the full details on the model we work with along with a derivation 
of all the $\wpri$ couplings in Appendix (App)~\ref{Coupl.APP}~and~\ref{coupling}. 
Here, we summarize some of the salient features of the various 
cases that we study in detail in the next two sections.
First of all, due to the extended EW gauge symmetry in the bulk, i.e.,
$SU(2)_L \times SU(2)_R \times U(1)_X$ which is motivated by
suppressing contributions to the $T$ parameter,
we see that there are two charged KK towers (one from each
$SU(2)$ group), before EWSB.
We will restrict to the 1st mode of each tower, denoting these states
by $W_{ L_1, \; R_1 }$, respectively. EWSB will mix these states and the
resulting 
mass eigenstates will be denoted by $\tilde{W}_{ L_1, \; R_1}$,
or to reduce clutter, simply as $W'_{L,\; R}$.

As explained above, these charged EW gauge bosons will decay mostly
into Higgs, including (longitudinal)  $W/Z$ and to top-bottom final states.
In the appendix, 
we define two cases for the top-bottom sector (corresponding to 
different representations for the top-bottom sector
under the $SU(2)_L\times SU(2)_R$ group) that we will consider in this work.
Here we summarize the main features of the two cases
(details
are given in the appendix)
\begin{itemize}
\item[] Case (i): 
$t_R$ has close-to-flat profile and
$(t,b)_L$ has a profile localized very close to the TeV brane in the bulk
\item[] Case (ii): 
vice versa, i.e., $(t,b)_L$ has close-to-flat profile and
$t_R$ has profile localized very close to the TeV brane in the bulk 
\end{itemize}
Roughly speaking, flavor precision tests tend to (strongly)
prefer case (ii), whereas EW precision tests have a (milder)
preference for case (i).

Since, as shown in the appendix,
the representations under $SU(2)_L \times SU(2)_R \times U(1)_X$
for these two cases are less than minimal, 
there are various``exotic'' 
fermion fields included in these
representations (cf Eqs.~(\ref{Q3Ldef.EQ})~and~(\ref{tRdef.EQ})) in 
addition to the SM fermions.
These exotic fermions can be looked for at the LHC but we will not 
consider them here;
instead we will restrict ourselves to SM final states.
However, in order to obtain realistic values, we will include these exotic 
decay channels in computing the BR's.


\section{$W^{ \prime }$ decays}
\label{decay}

\begin{figure}
\begin{center}
\scalebox{0.75}{\includegraphics[angle=270]{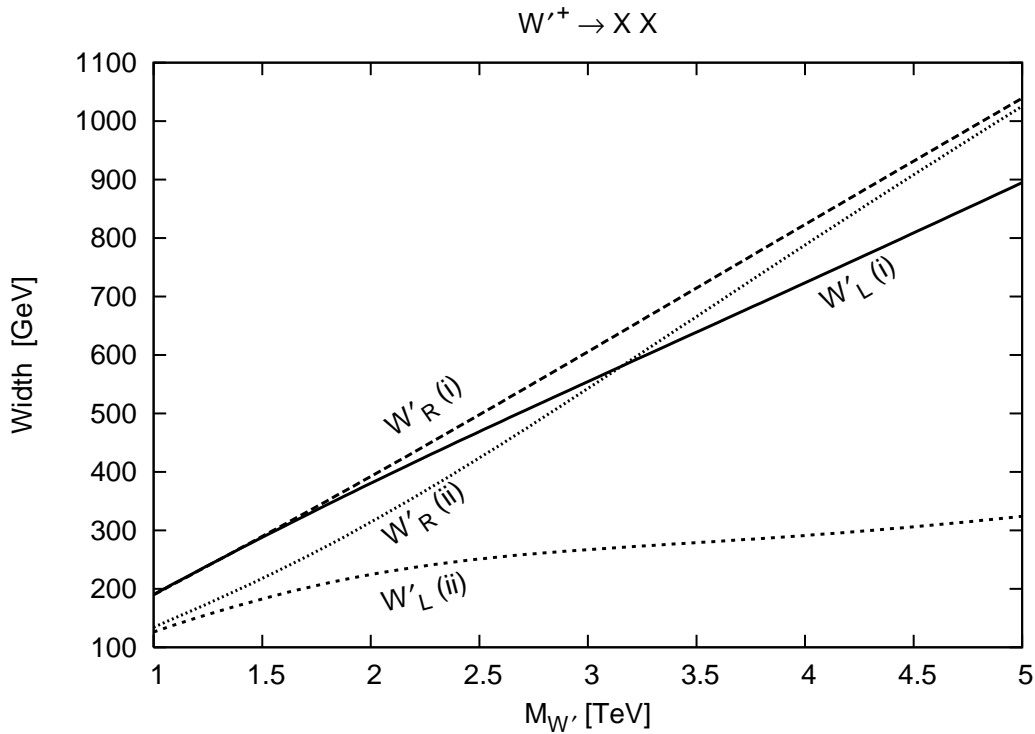}}
\caption{The total widths of $\wpLnt$ and $\wpRnt$ as a function of their masses for cases (i) and (ii). 
\label{totwidth.FIG} }
\end{center}
\end{figure}

%

The overlap integrals that dictate the 
$\wpri$ couplings to fermions in the four-dimensional effective theory 
are presented in Table~\ref{ovlap_ffG.TAB}. 
We then derive the $\wpri$ couplings to gauge bosons in the rest of App.~\ref{coupling}.
Armed with these couplings,
we are ready to embark on the phenomenology
of charged EW gauge bosons in this framework.
We have incorporated the $\wpri$ couplings shown in App.~\ref{coupling} 
into the Monte Carlo program CalcHEP~\cite{CalcHEP}, 
using which we present the results below. 
We will refer to the mass eigenstates $\wpLt$ and $\wpRt$ as just 
$\wpLnt$ and $\wpRnt$ respectively, i.e., we will always work with mass eigenstates
while studying the phenomenology.
In Fig.~\ref{totwidth.FIG} we show the total widths of the $\wpLnt$ and $\wpRnt$ into 2-body final states 
as a function of their masses for cases (i) and (ii). 
\begin{figure}
\begin{center}
\includegraphics[angle=0,width=0.49\textwidth]{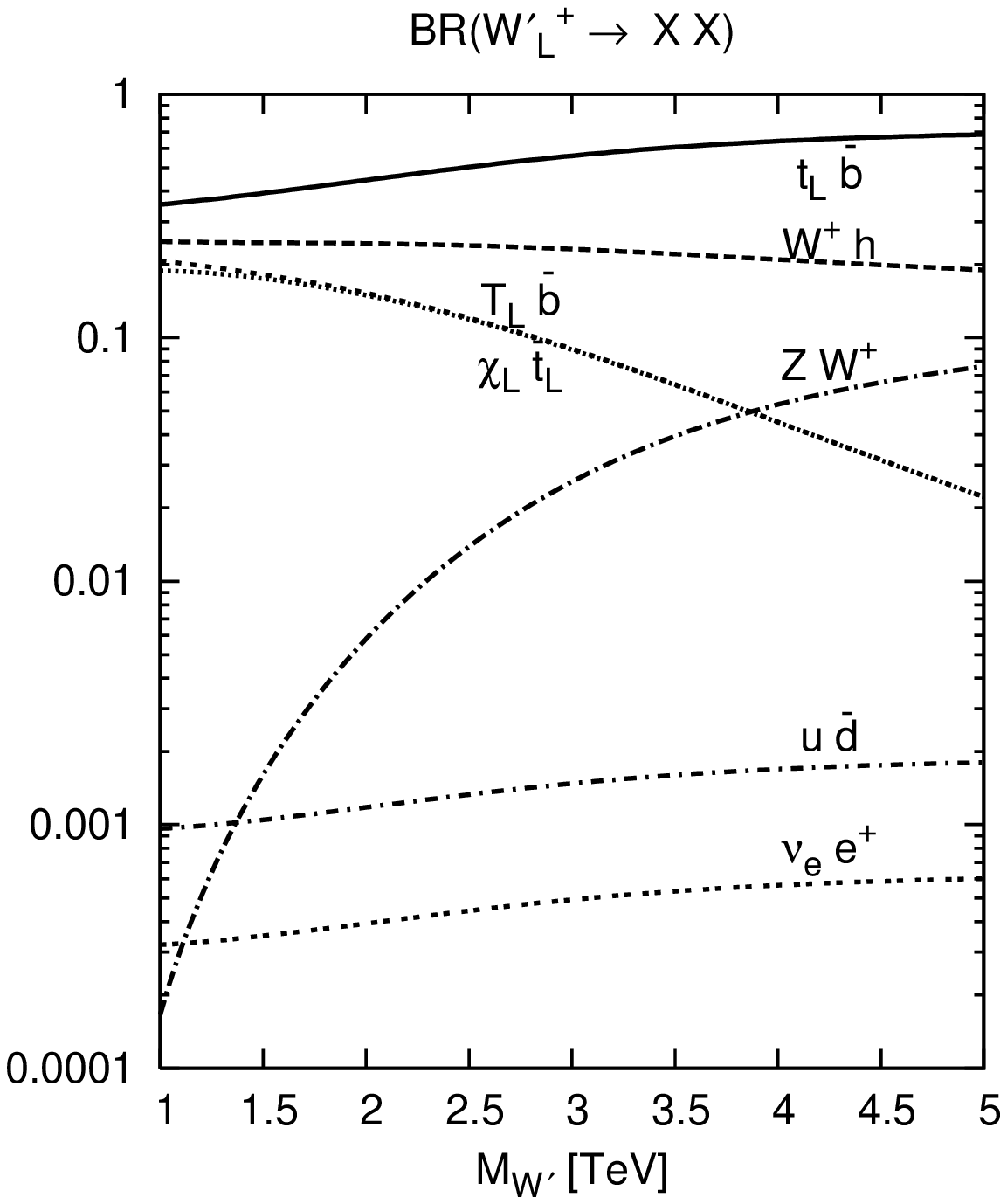}
\includegraphics[angle=0,width=0.49\textwidth]{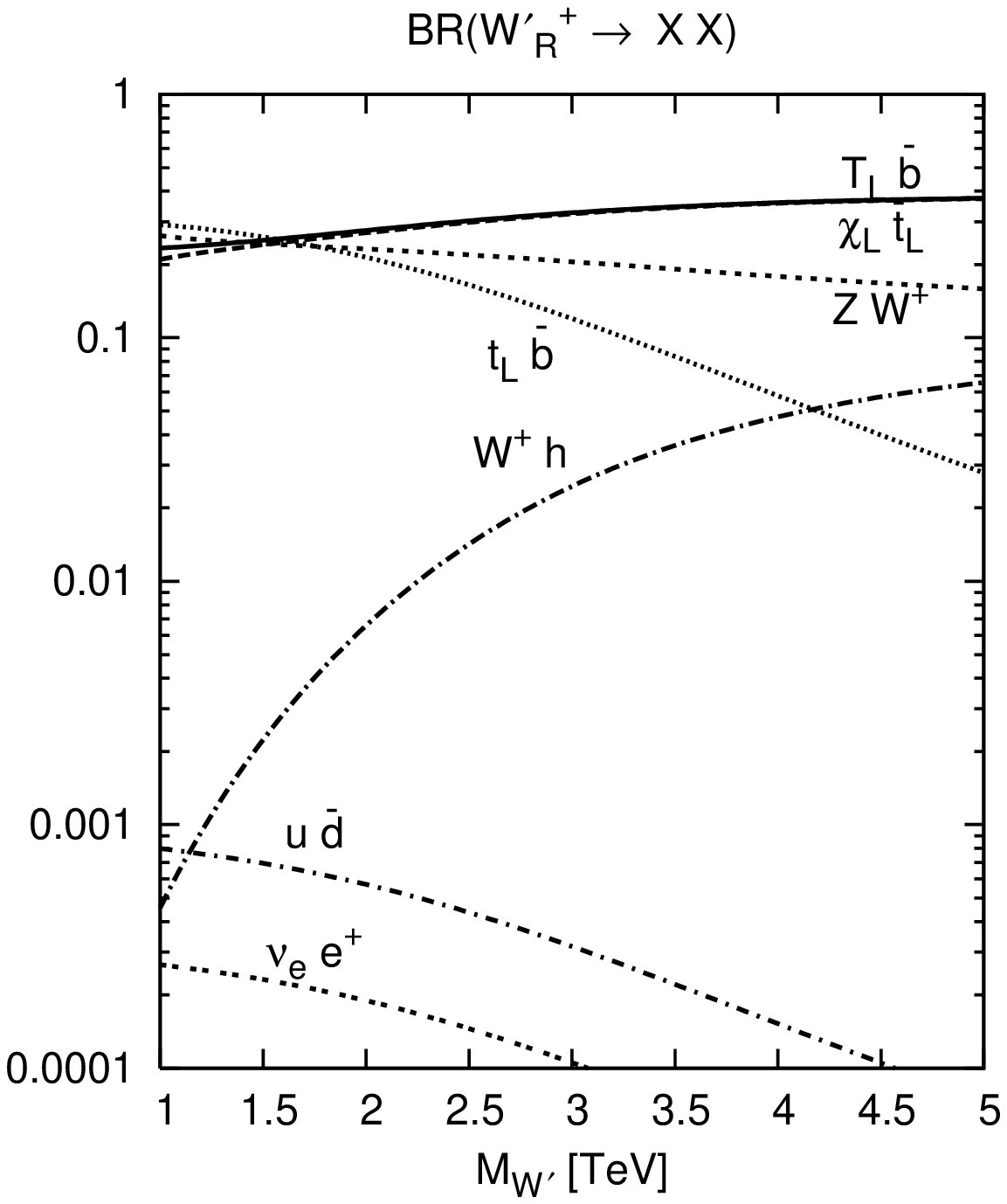}
\includegraphics[angle=0,width=0.49\textwidth]{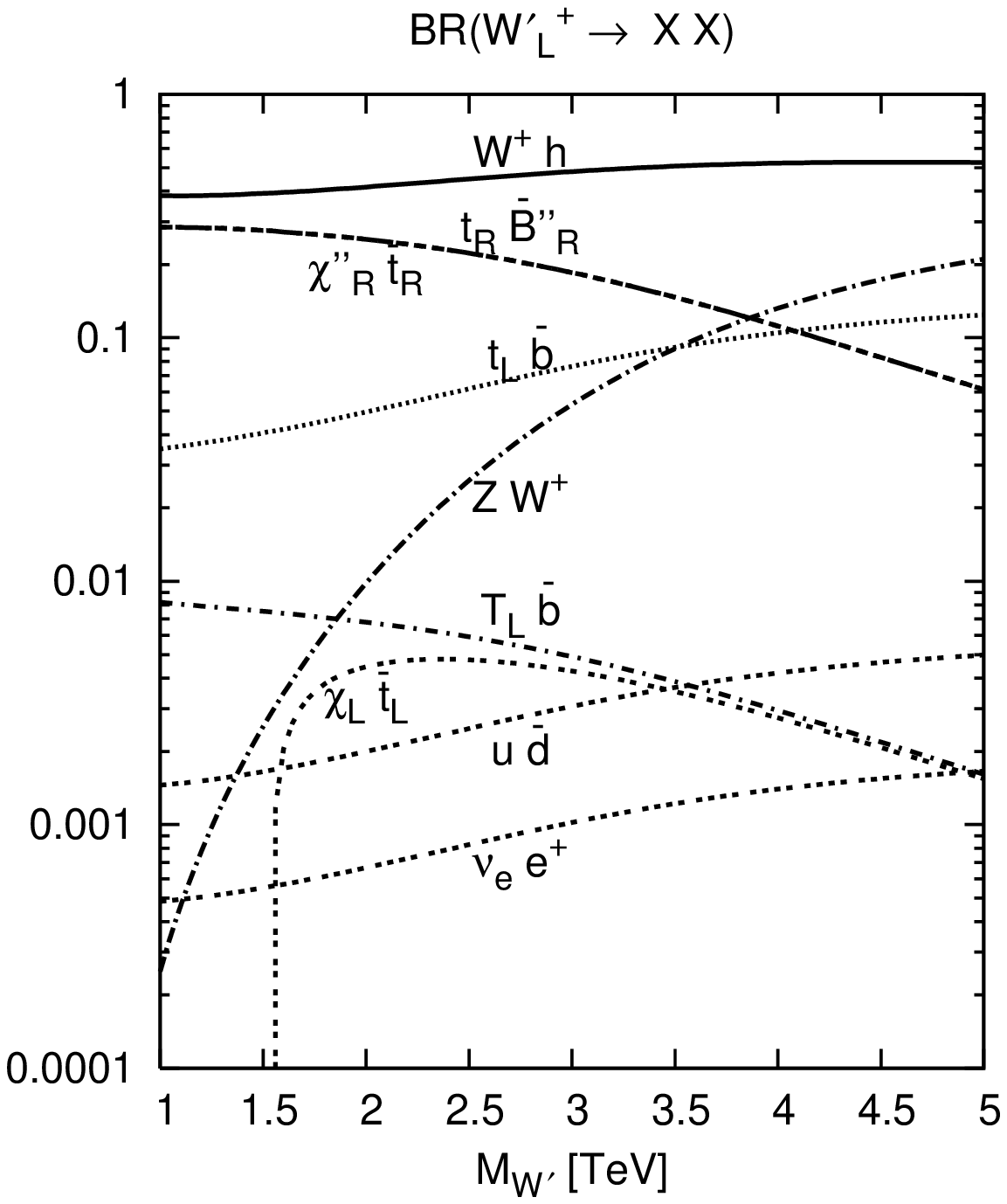}
\includegraphics[angle=0,width=0.49\textwidth]{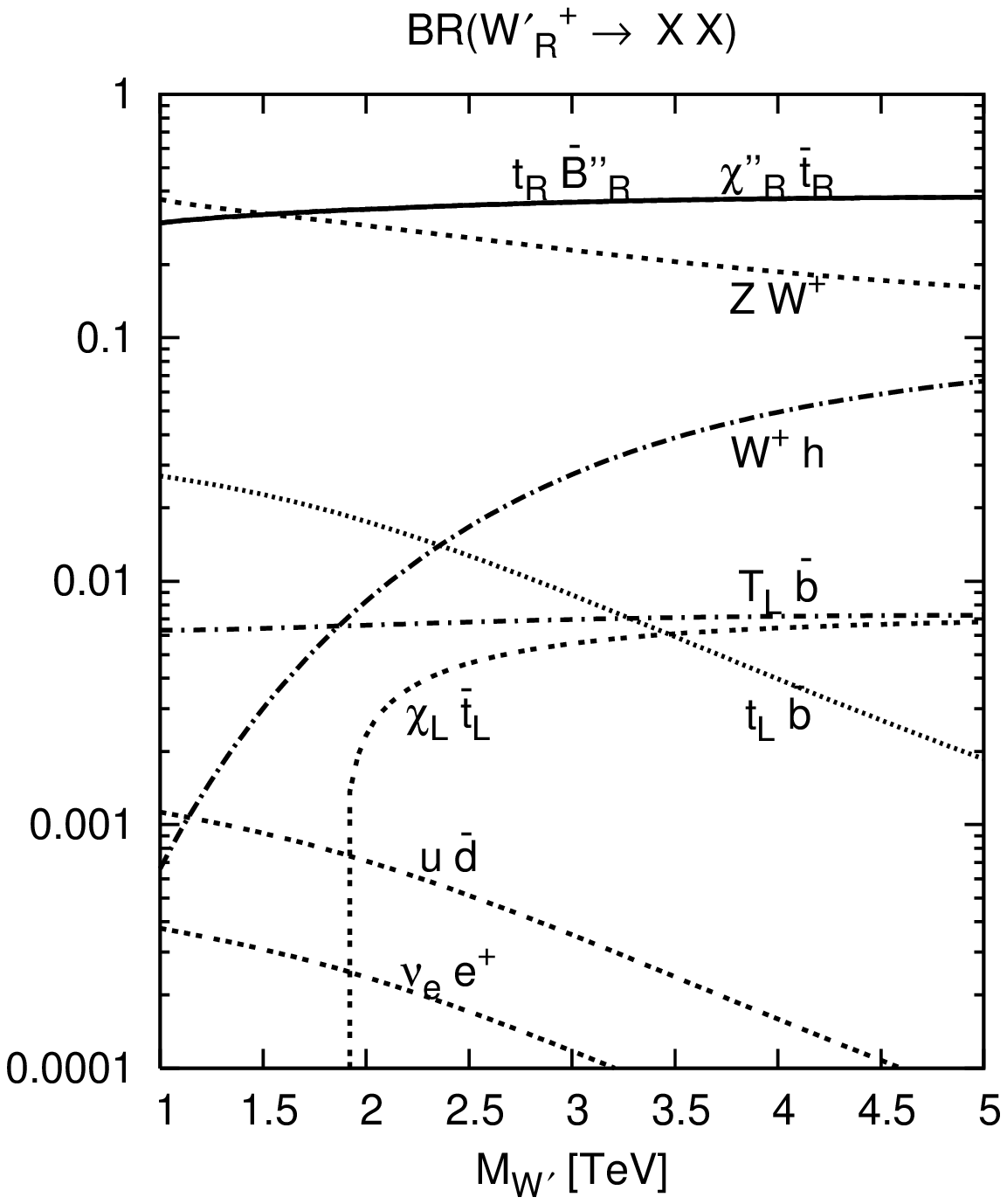}
\caption{The branching fractions of $\wpLnt$ (left) and $\wpRnt$ (right) as a function of their masses 
for case (i) (top panels), and case (ii) (bottom panels). In the bottom panel for case (ii),
note that the curves for $t_R \bar{B}^{\prime\prime}_R$ and $\chi^{\prime\prime}_R \bar{t}_R$ 
are on top of each other and cannot be individually differentiated.
\label{BRs.FIG} }
\end{center}
\end{figure} 
The total width increases monotonically with $\mwpri$ as expected, and is roughly
about $20\%$ or less of its mass, with it being appreciably smaller for 
the $W'_L$ in case (ii).  
This implies a typically weakly interacting particle
and a prompt decay, although the width can still be rather large for a high mass. 
A distinct feature
is that the total width of $\wpLnt$ in case (ii) is much smaller than all the others for
large $\mwpri$. This is due to the fact that  in case (ii) there is no direct coupling 
of the $\wpL$ to a third generation fermion whose wavefunction is peaked toward the TeV brane, 
unlike in the other cases. This is a direct result of the $SU(2)_L\otimes SU(2)_R$ 
quantum numbers
of the third generation fermions (cf Eqs.~(\ref{Q3Ldef.EQ})~and~(\ref{tRdef.EQ})). 
Note however that it can still be coupled to a TeV-brane peaked state via 
$\wpL\leftrightarrow \wpR$ mixing, but this would be suppressed by this mixing 
angle which is small for large $\mwpri$.

In Fig.~\ref{BRs.FIG} we show the branching fractions (BR) of the 
mass eigenstates $\wpLnt$ (left panel) and 
$\wpRnt$ (right panel) 
into various 2-body final states for cases (i) (top panels) and (ii) (bottom panels).
The largest branching fraction is to fermions peaked toward the TeV brane, 
which in case (i) is to $Q_L^3$ modes, while for $\wpRnt$ in case (ii) it is to the triplet containing $t_R$.
In contrast, for the $\wpLnt$ in case (ii) the largest BR is to $W h$ since there is no direct 
coupling to a third generation fermion with a TeV brane peaked profile.
For the $W'_R$ in case (ii), the $Z W$ final state is also available with a sizable BR. 
In case (ii) the $\chi_L \bar{t_L}$ is available only for $\mwpri > M_{\chi_L} + M_t$, 
and therefore exhibits a threshold behavior.

In this work we do not study the exotic particles in the final states, 
and we will focus on the $t \bar b$, $Z W$ and $W h$
final states in the rest of the paper. 
The BR into $\ell \nu$ final state is tiny, but due to its uniqueness and
for completeness we will briefly comment on this mode also. 
We will perform a detailed study of these final states considering their various decay modes
and obtain the LHC reach. 

\section{Charged Gauge Boson Signals at the LHC}
\label{signal}

\begin{figure}
\begin{center}
\scalebox{0.6}{\includegraphics[angle=0]{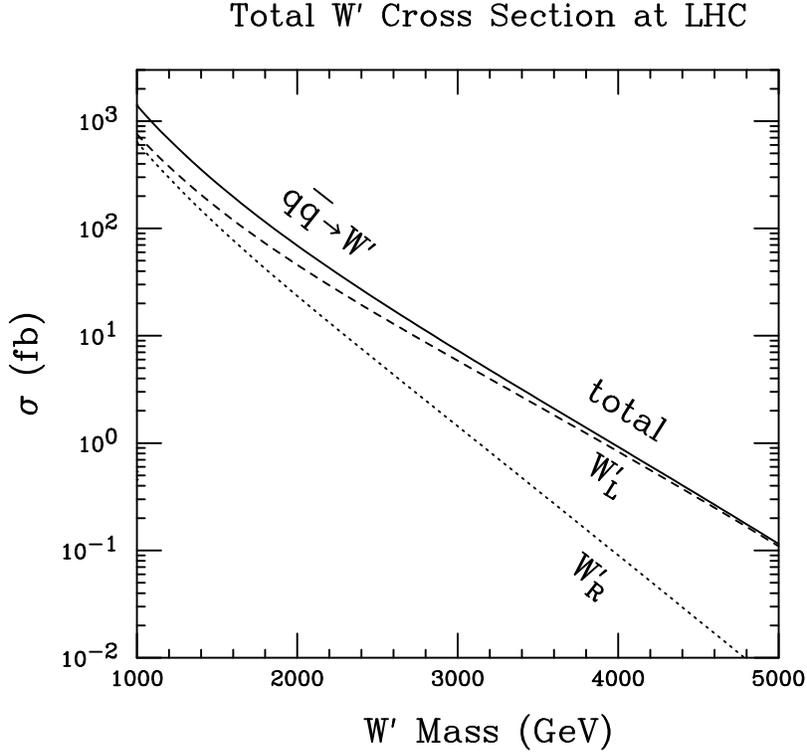}}
\caption{Total production cross-section for $W'$ versus its mass at the LHC.
\label{fig:tot} }
\end{center}
\end{figure} 

In this section we consider the production of the charged KK gauge bosons 
(generically denoted by  $\wpri$) and its decay into various SM final states
at the LHC. 
We first present in Fig.~\ref{fig:tot}
the total production cross-section for the $W'$ at the LHC versus its mass $M_{W'}$
via the Drell-Yan (DY) process.
We see that the cross-section can be $100 - 1$ fb for $M_{W'}=2-4$ TeV. 
The $\wpRnt$ coupling to light quarks is suppressed by the $W_{L_1}\leftrightarrow W_{R_1}$
mixing angle (cf Eq.~\ref{udWp.EQ}) and the rate is therefore smaller by a factor of 
$2-10$ in the mass range mentioned above.
In the following analyses, we coherently sum the $\wpLnt$ and $\wpRnt$ 
(the mass eigenstates) contributions.
There are other possible mechanisms for $W'$ production. One may consider the gauge
boson fusion $WZ \to W'$. However, the gauge-boson fusion channel, 
as first explored in \cite{Agashe:2007ki}, was found to be subleading. 
As mentioned earlier, we adopt the Monte Carlo package CalcHEP~\cite{CalcHEP} 
to obtain the numerical results in this section. We use the CTEQ6M for parton distribution
functions \cite{Pumplin:2002vw}.

\subsection{$t \bar b$ final state}
We first consider the production and decay channel 
\beq
pp\to {W^\prime}^+ \to t \bar{b}\quad {\rm with}\quad t\to b\  \ell \bar\nu\quad (\ell=e,\ \mu),
\label{eq:Wtb}
\eeq
where the leptonic decay modes of the top have been specified for the purpose of event
triggering and identification. 
The most distinctive feature of this signal is the large invariant mass of the $tb$ system 
near $M_{W'}$. Although the missing neutrino makes the event reconstruction less
trivial, one should be able to reconstruct the neutrino momentum fairly effectively
by demanding the two mass relations $M_W^2=(p_l+p_\nu)^2,\  m_t^2=(p_b+p_l+p_\nu)^2$.
We thus assume that the signal events are fully reconstructible.

\begin{figure}
\begin{center}
\scalebox{0.43}{\includegraphics[angle=270]{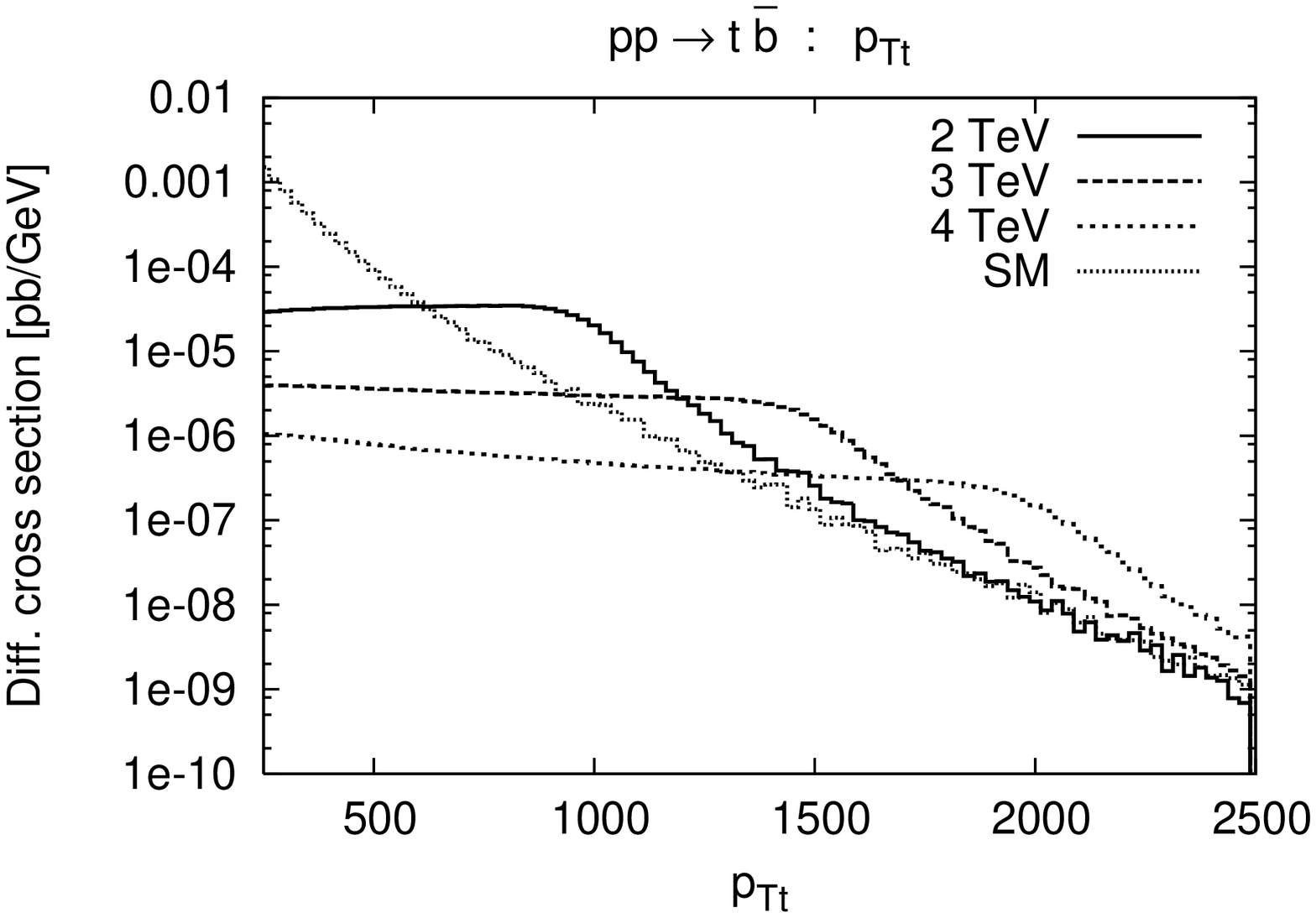}}
\scalebox{0.43}{\includegraphics[angle=270]{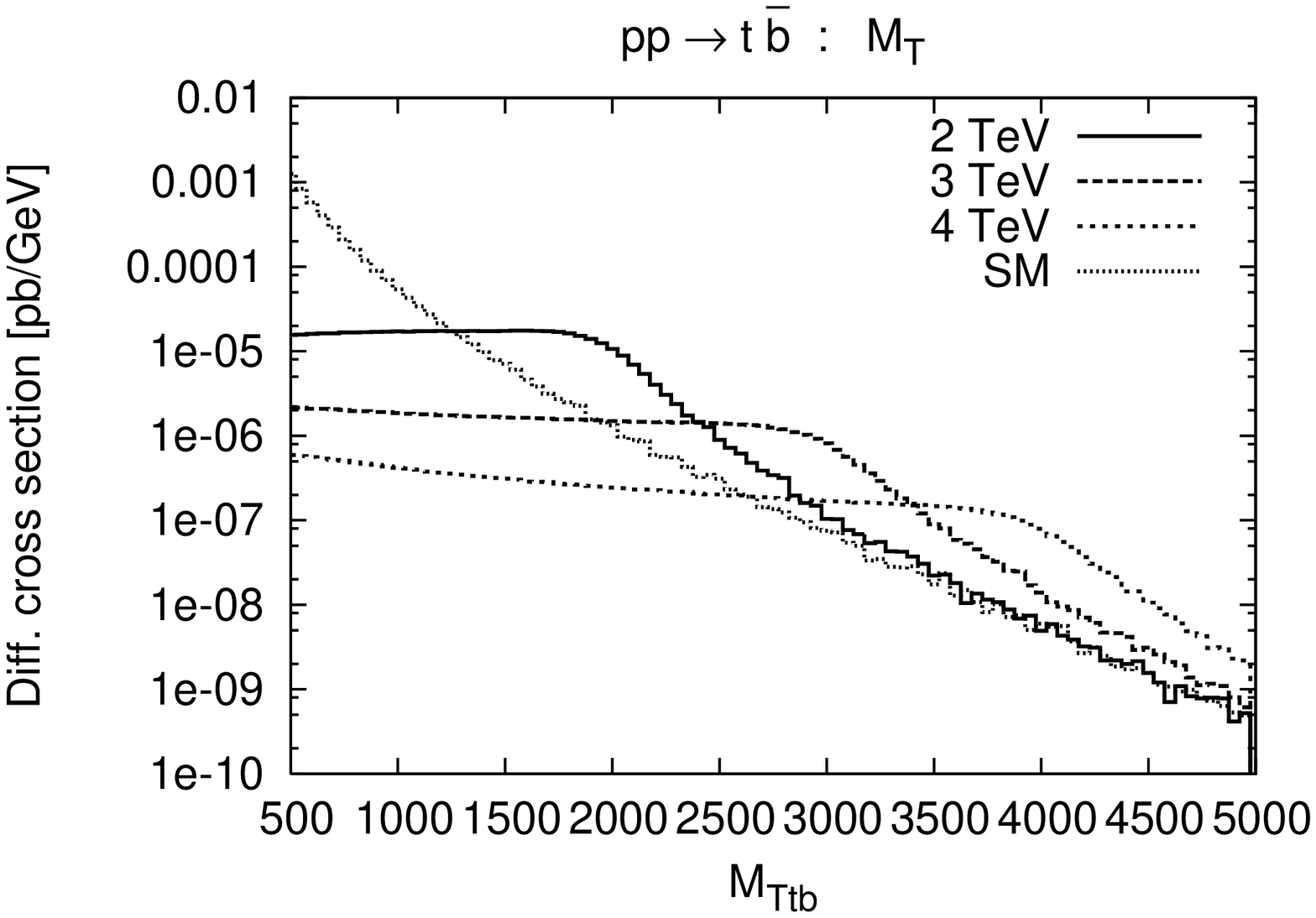}}
\caption{The (a) ${p_T}_t$, 
and (b) ${M_T}_{tb}$ 
differential distributions of the process 
$pp\to {W^\prime}^+ \to t \bar{b}$ for $M_{\wpri} = 2$, 3 and 4 TeV 
for case (i). These are after the cuts in Eq.~(\ref{eq:cutq}). 
Also shown are the SM single top background distributions. 
\label{pp2tb.FIG} }
\end{center}
\end{figure}

We select events with the basic acceptance cuts
\beq
|y_{W,b,\bar b}| < 3 ; \quad {p_T}_{W,b,\bar b} > 50~{\rm GeV}, 
\eeq
where the $y$'s are the rapidities.
For the signal events from a heavy $W'$ decay, further tightened cuts can help for the 
background suppression.  We thus impose the cuts on the top and $b$ quarks
\beq
\label{eq:cutq}
{p_T}_{t,b} > 200~{\rm GeV},\quad |y_{t,b}| < 3.
\eeq
In Fig.~\ref{pp2tb.FIG}, we present the differential distributions for the $t\bar b$ final state
with a variety of values of $\mwpri = 2$, $3$ and $4$~TeV, for (a) the transverse momentum
distribution, and (b) the transverse mass distribution. 
Also shown on the figures is the dominant source of irreducible background, 
the SM single top production $pp\to {W}^+ \to t \bar{b}$, seen as the continuum curves. 
We see the very promising prospects for observing the signal with suitably chosen cuts. 

\begin{figure}
\begin{center}
\scalebox{0.43}{\includegraphics[angle=270]{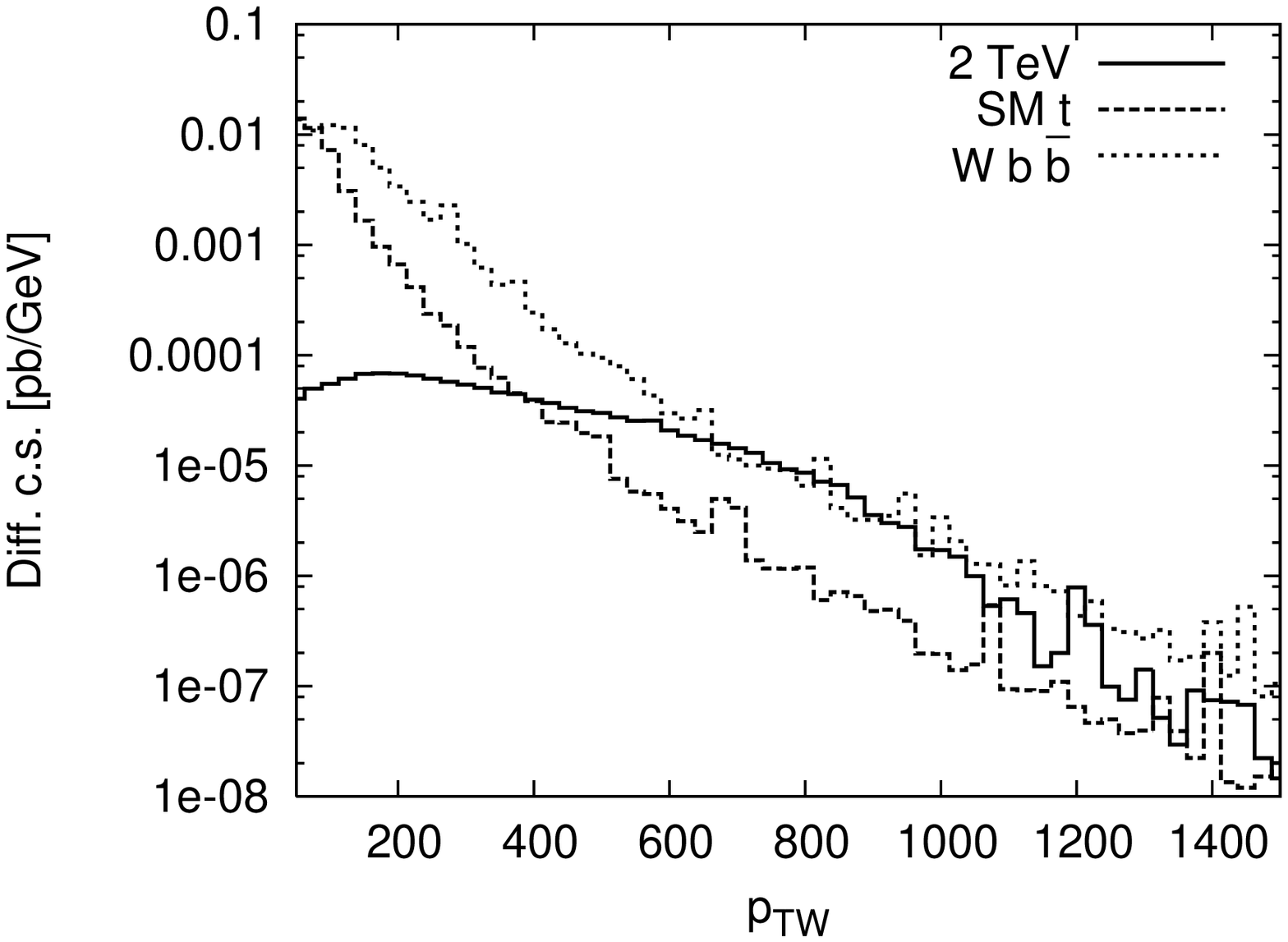}}
\scalebox{0.43}{\includegraphics[angle=270]{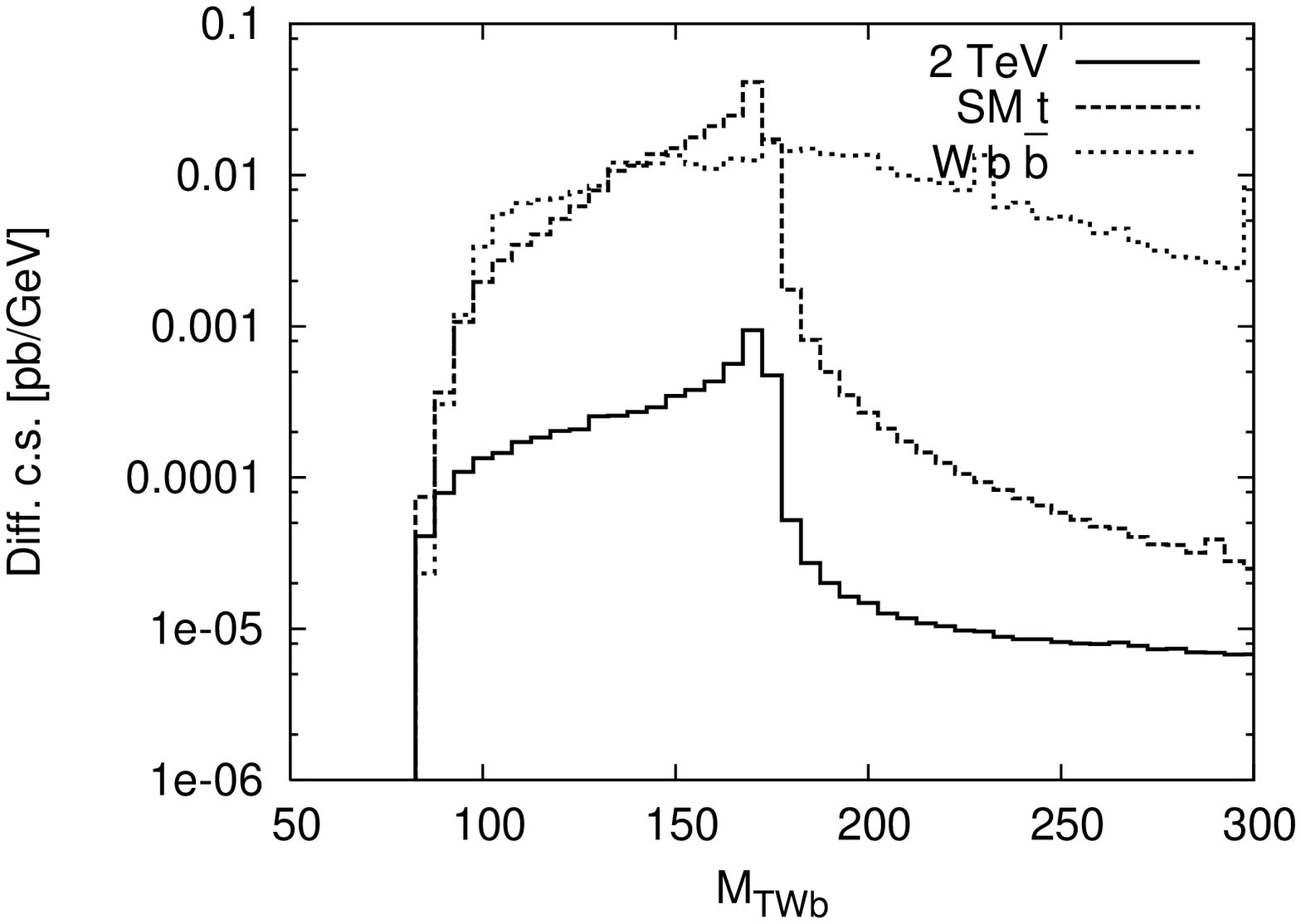}}
\scalebox{0.43}{\includegraphics[angle=270]{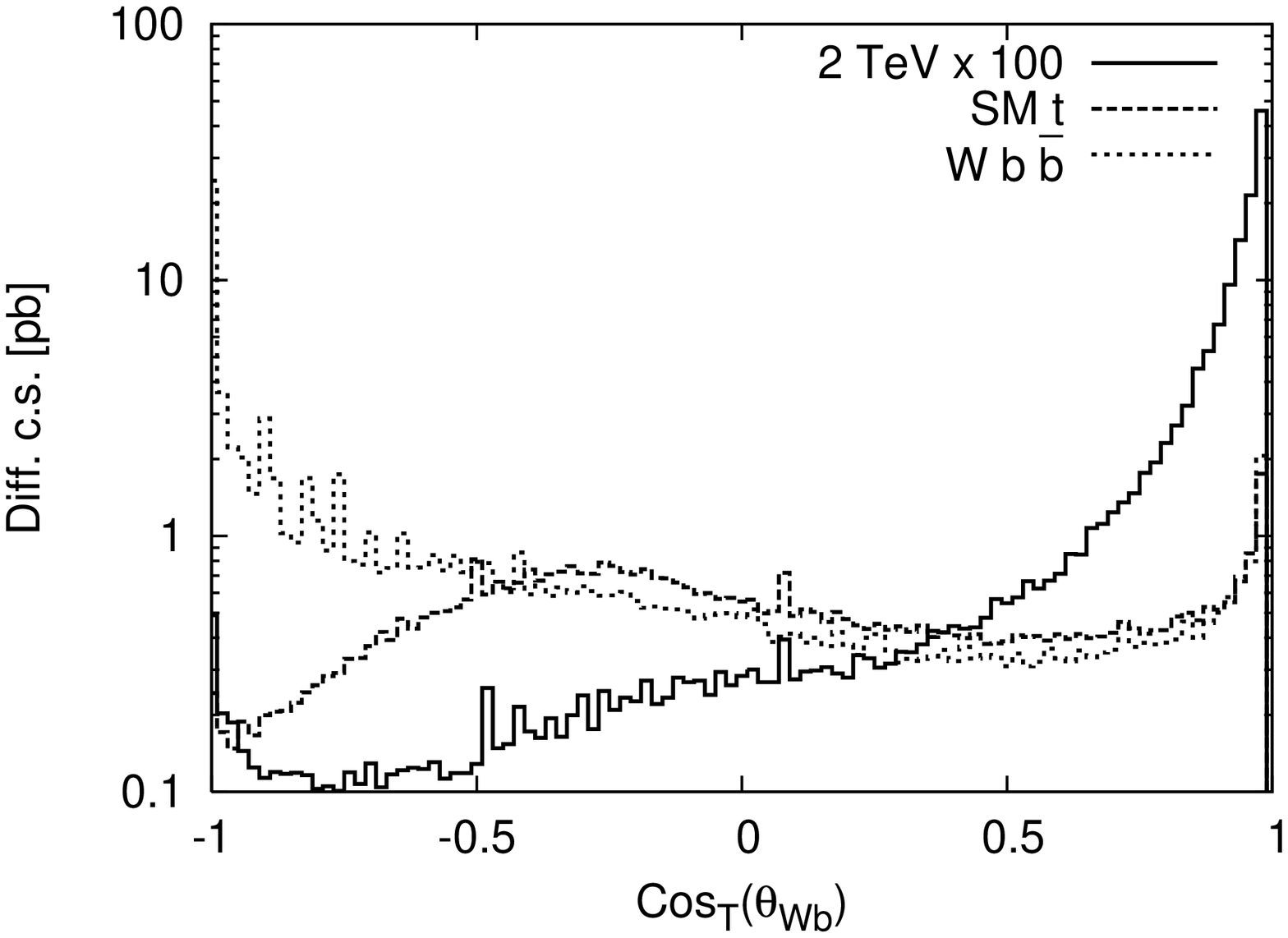}}
\scalebox{0.43}{\includegraphics[angle=270]{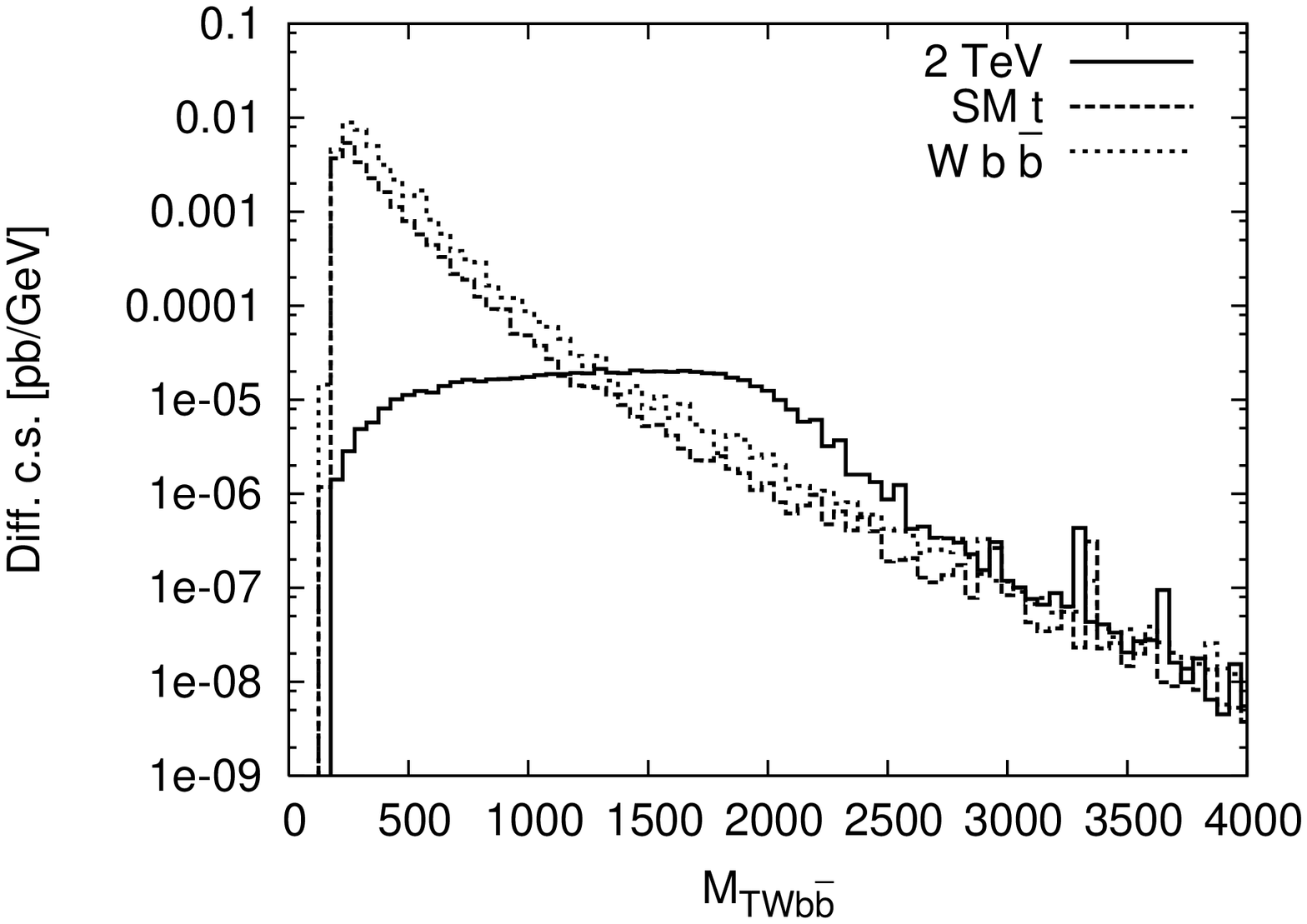}}
\caption{The (a) ${p_T}_W$, (b) ${M_T}_{Wb}$, (c) $\cos_T{\theta_{Wb}}$, 
and (d) ${M_T}_{Wb\bar b}$
differential distributions of the process $pp\to W^+ b \bar b$ for $M_{\wpri} = 2$ TeV, 
for case (i). Also shown are the irreducible backgrounds, the SM single top $t\bar b$
and $W b \bar b$. 
\label{pp2Wbb.FIG} }
\end{center}
\end{figure}

With the decay of the top (into $W b$), the SM $W b \bar b$ will be an additional source of background.
Since this background largely populates the low mass threshold region, we thus
form the following cluster transverse mass to help distinguish the signal from background
\bea
{M_T}_{Wb} &=& \left(\sqrt{{p_T}_W^2 + m_W^2} + {p_T}_b \right)^2 - \left|{{\bf p}_{T_W}} + {{{\bf p}}_{T_b}} \right|^2 \ , \nonumber \\
{M_T}_{Wb\bar b} &=& {p_T}_b + {p_T}_{\bar b} + \sqrt{{p_T}_W^2 + m_W^2}  .
\eea
We show the representative kinematical distributions 
for the $W b \bar b$ final states in Fig.~\ref{pp2Wbb.FIG},
with a $\mwpri = 2$~TeV signal (solid curves) for illustration. The SM backgrounds 
of $t\bar b$ (dashed) and $Wb\bar b$ (dotted) are also shown for comparison.
Figure \ref{pp2Wbb.FIG}(a) presents the transverse momentum distribution for the $W$. 
In Fig.~\ref{pp2Wbb.FIG}(b), we show the distribution of ${M_T}_{Wb}$, 
the cluster transverse mass of the $W$ with the nearer of the two b-jets. The top-quark
mass reconstruction is visible for those events with a real top in the final state.
We show in Fig.~\ref{pp2Wbb.FIG}(c) the distribution of $\cos_T{\theta_{Wb}}$, 
the cosine of the 
angle in the transverse plane between the $W$ and the nearer of the two b-jets. 
Due to the large boost of the top quark from the $W'$ decay for the signal, 
the opening angle obviously is rather small as seen by the solid curve. 
In Fig.~\ref{pp2Wbb.FIG}(d), we show the distribution of the full transverse mass 
of the $W b \bar b$ system. It is encouraging to see a possible separation of
the signal from the backgrounds.

Since  there is  only one missing neutrino, the kinematical 
variables can be fully reconstructed in the event as discussed earlier
by demanding the mass reconstruction of $M_W,\ m_t$. 
Alternatively, since the $W$ is produced with a large boost, 
the neutrino will be considerably collimated with the  charged-lepton. 
If one makes the 
assumption\footnote{Practically, 
$p_T(\nu)=\etmiss,\  p_L(\nu)=p_L(\ell) \times{\etmiss\over p_T(\ell)} $.}
$\vec p_\nu \approx \kappa \vec p_\ell$, 
the neutrino 4-momentum can be approximately determined
and the full $M_{Wb\bar b}$ can be formed. We find that doing so gives a narrower signal 
invariant-mass peak,
but also raises the background in the region of interest, resulting in a marginal improvement in
significance; we therefore do not pursue either of these ideas here.

\begin{figure}
\begin{center}
\scalebox{0.53}{\includegraphics[angle=270]{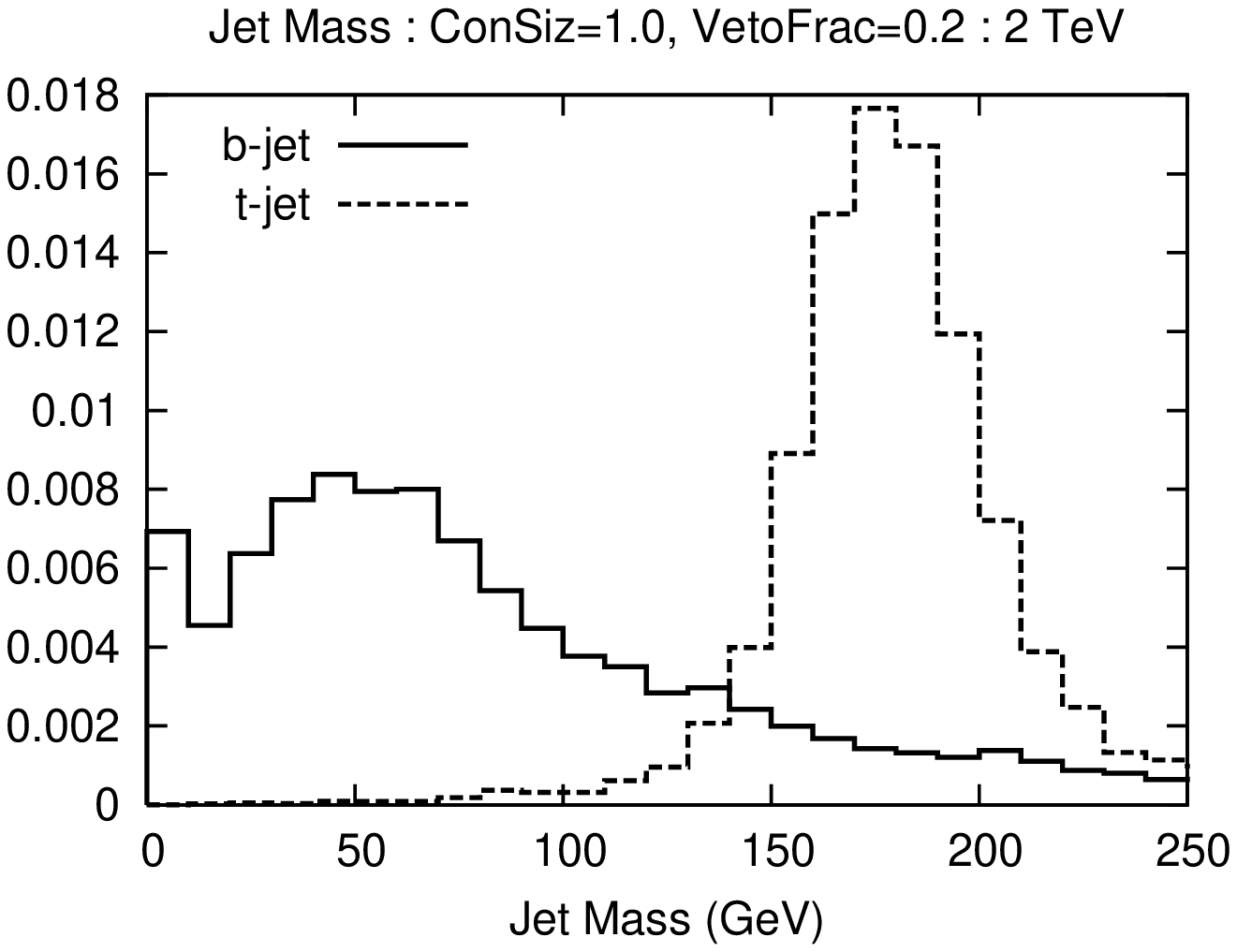}}
\scalebox{0.53}{\includegraphics[angle=270]{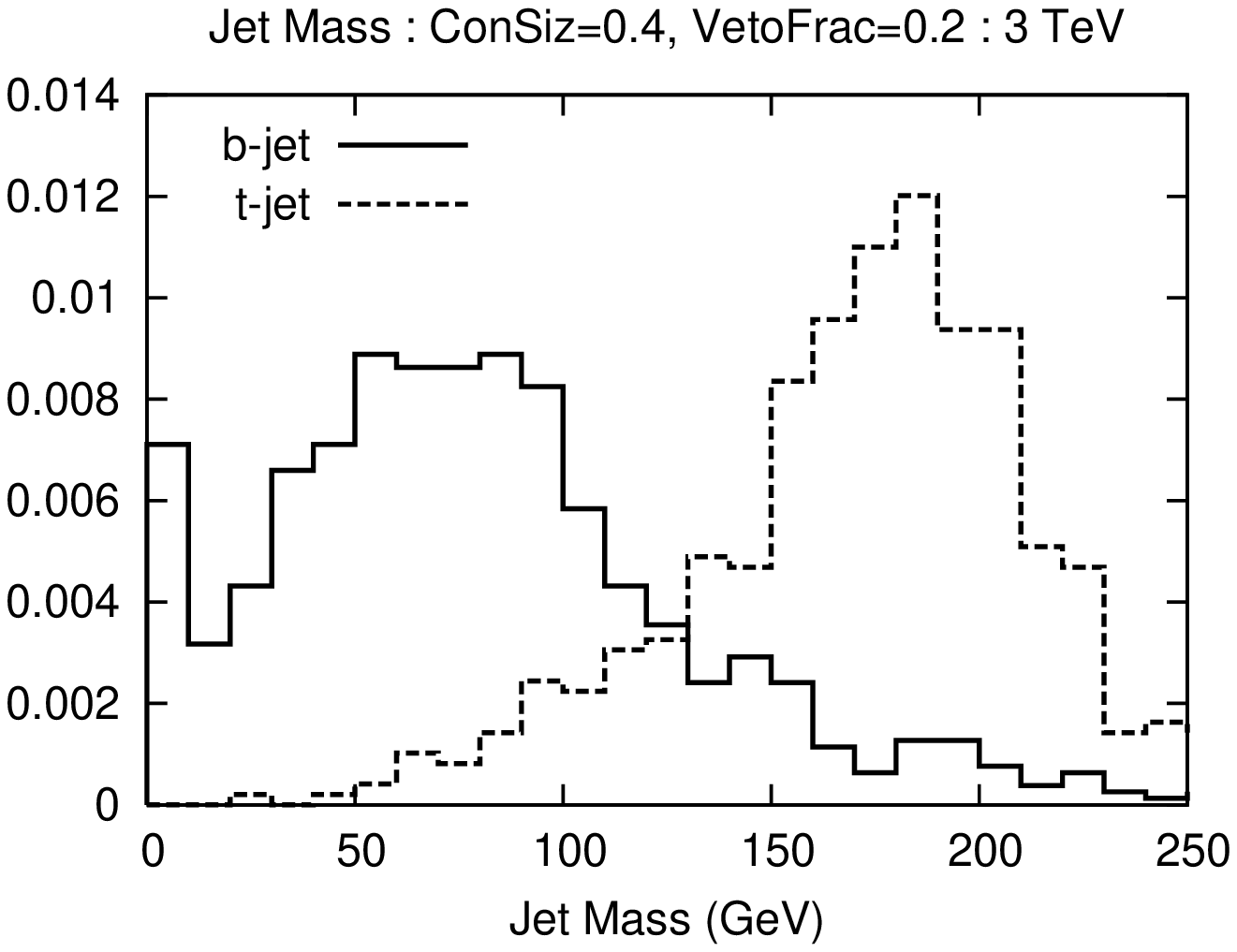}}
\caption{The jet-mass distributions for top and bottom jets.
These are after basic, $Wb$ and invariant-mass cuts. 
On the left is for 2~TeV with cone-size 1.0, and,
on the right for 3~TeV with cone-size 0.4. Both are with veto-fraction of 0.2.
\label{jm.FIG} }
\end{center}
\end{figure}

The distributions in Fig.~\ref{pp2Wbb.FIG} motivate us to consider the following cuts: 

\noindent
\underline{ ${M_T}_{Wb}$ cut:} 
$100 < {M_T}_{Wb} < 190~{\rm GeV}$, since we expect for the signal this should reconstruct to the 
parent $m_{\rm top}$, as can be seen in Fig.~\ref{pp2Wbb.FIG}(b). 
Notice that since for single top this also reconstructs
to $M_{\rm top}$, this variable does not discriminate between the signal and this 
source of background. \\
\underline{$Wb$ angle cut}: $\cos_T{\theta_{Wb}} > 0.5$, motivated from (c), where we see that 
for the signal, the $W b$ opening angle is fairly small owing to the large boost of the parent top. \\
\underline{2 $b$-tags}: We demand that there be two tagged $b$'s in the event.
We adopt a $b$-tagging efficiency $\eta_b = 0.4$ \cite{btagRef}. 
With $b$-tagging parameters optimized for low ${p_T}_b$ the light-quark rejection ratio
(for $j=u,d,s,g$) is roughly $R_j = 20$~\cite{btagRef},
where $1/R_j$ is the probability of mistaking a light-jet for a b-jet. 
We believe this is likely to be improved with tagging techniques optimized for high ${p_T}_b$,
and since our light-quark jet background is significant, we anticipate such improvements 
and use $R_j = 40$.
We use a charm quark rejection factor $R_c = 5$. \\
\underline{${M_T}_{WbB}$ cut}: $1500 < {M_T}_{Wb\bar b} < 2500~{\rm GeV}$ (for $\mwpri = 2$~TeV), and,  
$2400 < {M_T}_{Wb\bar b} < 3600~{\rm GeV}$ (for $\mwpri = 3$~TeV), which is motivated by the resonant 
feature of the signal as seen clearly in (d), and results in the background being very effectively 
suppressed after this cut. \\
\underline{Jet-mass cut}: $t\bar t$ production can become a source of background
since a top can fake a $b$-jet, for instance when 
the hadronic decay products of a boosted top are sufficiently collimated
that it can be confused for a $b$-jet. 
The two main sources of a top pair are the SM QCD production, and 
the KK-gluon production which dominantly decays to this channel. Both can be significantly 
larger than the signal, and the latter is especially problematic since it is resonant in the same
invariant mass region as the signal. However, the jet-mass variable can be used to discriminate 
between a $b$-jet and a boosted hadronic top, 
with the distributions expected to peak at $m_b$ and $m_t$ respectively. 
In order to obtain a rough estimate of 
the separation achievable, we have used Pythia v6.411~\cite{Sjostrand:2006za} 
to shower a bottom and a hadronically decayed top, followed by smearing the daughter particles
energy by $80$\,\%/$\sqrt{E}$, and $\eta$ and $\phi$ by $0.05$
to mimic the finite resolutions of the detector.\footnote{We are grateful to Frank Paige
for many discussions on jet-mass issues.
This variable was also explored in Ref.~\cite{Agashe:2007ki} 
for the jet-mass of the $W$.
Related issues, including using sub-structure of
jets to reduce QCD backgrounds, have also been discussed in 
Refs.\cite{Lillie:2007yh, Thaler:2008ju, Kaplan:2008ie,jmassRefs}.
}    
In general, a larger cone-size will include more of the radiation and results
in a narrower distribution for $t$-jets but at the expense of moving the 
$b$-jet peak to larger values. 
Also, since a $b$-jet is expected to be more collimated than a $t$-jet, we also demand that 
$80$\% of the $p_T$ be contained within the cone (veto-fraction of $0.2$). 
We show in Fig.~\ref{jm.FIG}
the resulting jet-mass distributions for the 2~TeV (left) and 3~TeV (right) cases. 
For the 2~TeV case, we find a cone-size of 1.0 to result in adequate separation, 
and a jet-mass cut $M_j < 75$~GeV with a veto-fraction of 0.2 retains
46\% of b-jets and 0.38\% of $t$-jets.
For the 3~TeV case, the decay products are more collimated and we therefore pick
a smaller cone-size, namely 0.4, and again with a veto-fraction of 0.2. We find that, 
with the cut $M_j < 100$~GeV (larger than the previous case in order to keep
more of the already small signal events), 
57\% of $b$-jets and 2.5\% of $t$-jets are retained.

\begin{table}
\begin{center}
\caption{The cross-sections (in fb) for the signal process 
$pp\to \wpri \to t b \to W \bar b b \to \ell \nu \bar b b$ for case (i),
and SM background, with the cuts applied successively. 
Cross-sections are shown for $M_\wpri = 2$ and $3$~TeV, with the number of events and significance for  ${\cal L} = 100$ and $300$ ${\rm fb^{-1}}$, respectively. 
In the last two columns we show the significance without and with $t \bar t$
as a source of background, with the significance including the latter 
shown in parenthesis. 
\label{pp2lnbB.TAB}}
\begin{tabular}{|c|c|c|c|c|c|c|c|c|}
\hline 
2 TeV,\  100 ${\rm fb^{-1}}$
&
Basic&
$Wb$ cuts&
b-tag&
$M_{TWbb}$ &
$M_j$&
\# Evt&
$S/B$&
$S/\sqrt{B}$\tabularnewline
\hline
\hline 
Case (i)&
$8.9$&
$7$&
$1.1$&
$0.44$&
$0.2$&
$20$&
$2.5$ $(1.4)$&
$7$ $(5.3)$\tabularnewline
\hline 
SM top&
$1400$&
$370$&
$60$&
$0.09$&
$0.04$&
$4$&
&
\tabularnewline
\hline 
SM $W\, b\,\bar{b}$&
$520$&
$66$&
$11$&
$9\times10^{-3}$&
$4\times10^{-3}$&
$0.4$&
&
\tabularnewline
\hline
SM $W\, b\, j$&
$9\times10^{3}$&
$2\times10^{3}$&
$20$&
$0.04$&
$0.02$&
$2$&
&
\tabularnewline
\hline
SM $W\, c\, j$&
$4\times10^{3}$&
$700$&
$4$&
$10^{-3}$&
$0.5\times10^{-3}$&
$0.05$&
&
\tabularnewline
\hline
SM $W\, j\, j$&
$2\times10^{5}$&
$2\times10^{4}$&
$13$&
$0.03$&
$0.01$&
$1$&
&
\tabularnewline
\hline
\hline 
SM $t\bar{t}$&
$4\times10^{4}$&
$10^{4}$&
$2\times10^{3}$&
$4.5$&
$0.02$&
$2$&
&
\tabularnewline
\hline
$G^{(1)}$ $t\bar{t}$ (i)&
$250$&
$190$&
$30$&
$10$&
$0.04$&
$4$&
&
\tabularnewline
\hline
\end{tabular}

\medskip{}

\begin{tabular}{|c|c|c|c|c|c|c|c|c|}
\hline 
3 TeV &
Basic&
$Wb$ cuts&
b-tag&
$M_{TWbb}$&
$M_j$&
\# Evt&
$S/B$&
CL\tabularnewline
300 ${\rm fb^{-1}}$
& & & & & & & & Poisson \tabularnewline\hline
\hline 
Case (i)&
$1.5$&
$1.1$&
$0.18$&
$0.04$&
$0.02$&
$7$&
$5.8$ $(0.9)$&
$0.995\ (0.95)$\tabularnewline
\hline 
SM top&
$1400$&
$370$&
$60$&
$4\times10^{-3}$&
$2\times10^{-3}$&
$0.6$&
&
\tabularnewline
\hline 
SM $W\, b\,\bar{b}$&
$520$&
$66$&
$11$&
$4\times10^{-4}$&
$2.3\times10^{-4}$&
$0.07$&
&
\tabularnewline
\hline
SM $W\, b\, j$&
$9\times10^{3}$&
$2\times10^{3}$&
$20$&
$10^{-3}$&
$0.5\times10^{-3}$&
$0.2$&
&
\tabularnewline
\hline
SM $W\, c\, j$&
$4\times10^{3}$&
$700$&
$4$&
$10^{-4}$&
$0.5\times10^{-4}$&
$0.02$&
&
\tabularnewline
\hline
SM $W\, j\, j$&
$2\times10^{5}$&
$2\times10^{4}$&
$13$&
$2\times10^{-3}$&
$10^{-3}$&
$0.3$&
&
\tabularnewline
\hline
\hline 
SM $t\bar{t}$&
$4\times10^{4}$&
$10^{4}$&
$2\times10^{3}$&
$0.21$&
$5.3\times10^{-3}$&
$1.6$&
&
\tabularnewline
\hline
$G^{(1)}$ $t\bar{t}$ (i)&
$32$&
$24$&
$4$&
$0.64$&
$0.02$&
$5$&
&
\tabularnewline
\hline
\end{tabular}
\end{center}
\end{table}

In Table~\ref{pp2lnbB.TAB} we present the signal and background cross-sections (in fb) for the process 
$pp\to W b\bar b \to \ell\nu b \bar b$ for case (i).  We include in the signal 
both ${W^\prime}^+$ and ${W^\prime}^-$,  
and we find that the latter cross-section is about 
a third of the former, stemming from the difference in the PDF's of more $u$ quarks than $d$ quarks
in a proton. We count for both $\ell = e,\mu$. 
Due to the large boost of the parent top, the lepton may not have
a large isolation with respect to the b-jet, but we will assume that this will
not result in too large a loss of efficiency. 
Ways to deal with this has been discussed in Refs.\cite{kkgluon, Thaler:2008ju}. 
%
%
The entry labeled as ``SM top", in addition to the 
SM $W^\pm$-exchange 
single-top process, 
also includes the $W$-glue fusion process containing an 
extra jet that we use to 
veto events
with ${p_T}_j > 20$~GeV in the central region.  
In the last two columns we show the significance without and with $t \bar t$
as a source of background, with the significance including the latter 
shown in parenthesis. The $G^{(1)}$ (KK gluon) is taken to be degenerate with $\wpri$.
From the table we see that we need ${\cal L}=100~{\rm fb}^{-1}$ 
(${\cal L}=300~{\rm fb}^{-1}$) for a $2~$TeV ($3~$TeV) $\wpri$, where we have estimated 
the signal significance by $S/\sqrt{B}$ in Gaussian distribution 
for large event sample, but by Confidence Level (CL)
in Poisson statistics for small even sample. 
Although the $S/B$ is good for heavier masses, the signal will still be limited by 
statistics.

\begin{table}
\begin{center}
\caption{The cross-sections (in fb) for the signal process 
$pp\to \wpri \to t b \to W \bar b b \to \ell \nu \bar b b$ for case (ii),
and SM background, with the cuts applied successively. 
Cross-sections are shown for $M_\wpri = 2$~TeV, with the number of events and significance for 
${\cal L} = 1000~{\rm fb^{-1}}$. 
In the last two columns we show the significance without and with $t \bar t$
as a source of background, with the significance including the latter 
shown in parenthesis. 
\label{pp2lnbB_c2.TAB}}
\begin{tabular}{|c|c|c|c|c|c|c|c|c|}
\hline 
2 TeV&
Basic&
$Wb$ cuts&
b-tag&
$M_{TWbb}$ &
$M_j$&
\# Evt&
$S/B$&
$S/\sqrt{B}$\tabularnewline
\hline
\hline 
Case (ii)&
$0.75$&
$0.6$&
$0.1$&
$0.05$&
$0.03$&
$30$&
$0.38\,(0.2)$&
$3.4\,(2.5)$\tabularnewline
\hline 
SM top&
$1400$&
$370$&
$60$&
$0.09$&
$0.04$&
$40$&
&
\tabularnewline
\hline 
SM $Wj_{1}j_{2}$&
$2.1\times10^{5}$&
$2.2\times10^{4}$&
$48$&
$0.08$&
$0.04$&
$40$&
&
\tabularnewline
\hline
\hline 
SM $t\bar{t}$&
$4\times10^{4}$&
$10^{4}$&
$2\times10^{3}$&
$4.5$&
$0.02$&
$20$&
&
\tabularnewline
\hline
$G^{(1)}$ $t\bar{t}$ (ii)&
$210$&
$180$&
$29$&
$13$&
$0.05$&
$50$&
&
\tabularnewline
\hline
\end{tabular}
\end{center}
\end{table}

In Table~\ref{pp2lnbB_c2.TAB} we present the signal and background cross-sections (in fb) for the process 
$pp\to W b\bar b \to \ell\nu b \bar b$ for case (ii).
Rather than repeating, in Table~\ref{pp2lnbB_c2.TAB} we have combined, in SM $W j_1 j_2$, the 
SM $W b \bar b$, $W b j$, $W c j$ and $W j j$ channels shown separately in 
Table~\ref{pp2lnbB.TAB}. 
Compared to case (i), as expected, the cross-section is lower in case (ii) since the 
$t_L, b_L$ profiles are not peaked near the TeV brane (and hence
do not have as large coupling to $\wpri$) as in the former case, but rather
it is the $t_R$ which either does not couple to $\wpri$ if it is an $SU(2)_R$ singlet, 
or would couple to it only associated with an exotic fermion if it is a triplet.
We find that we need a much higher luminosity, namely, $1000~{\rm fb}^{-1}$ for a 
$2~$TeV $\wpri$. In this case, we expect the other channels (to be discussed in the
following) to have
better reach.

\subsection{$Z W$ final state}
As for the process
\beq
pp \to W' \to ZW,
\eeq
we consider the gauge boson decay modes separately. In order to effectively
reconstruct the final state, we do not pursue the missing decay channel $Z\to \nu\nu$.
Although {\em simultaneous} hadronic decays of $Z$ and $W$ have the largest branching fraction, 
the multiple jet background from QCD would be overwhelming. 
We therefore do not pursue this mode in our study.

\begin{figure}
\begin{center}
\scalebox{0.43}{\includegraphics[angle=270]{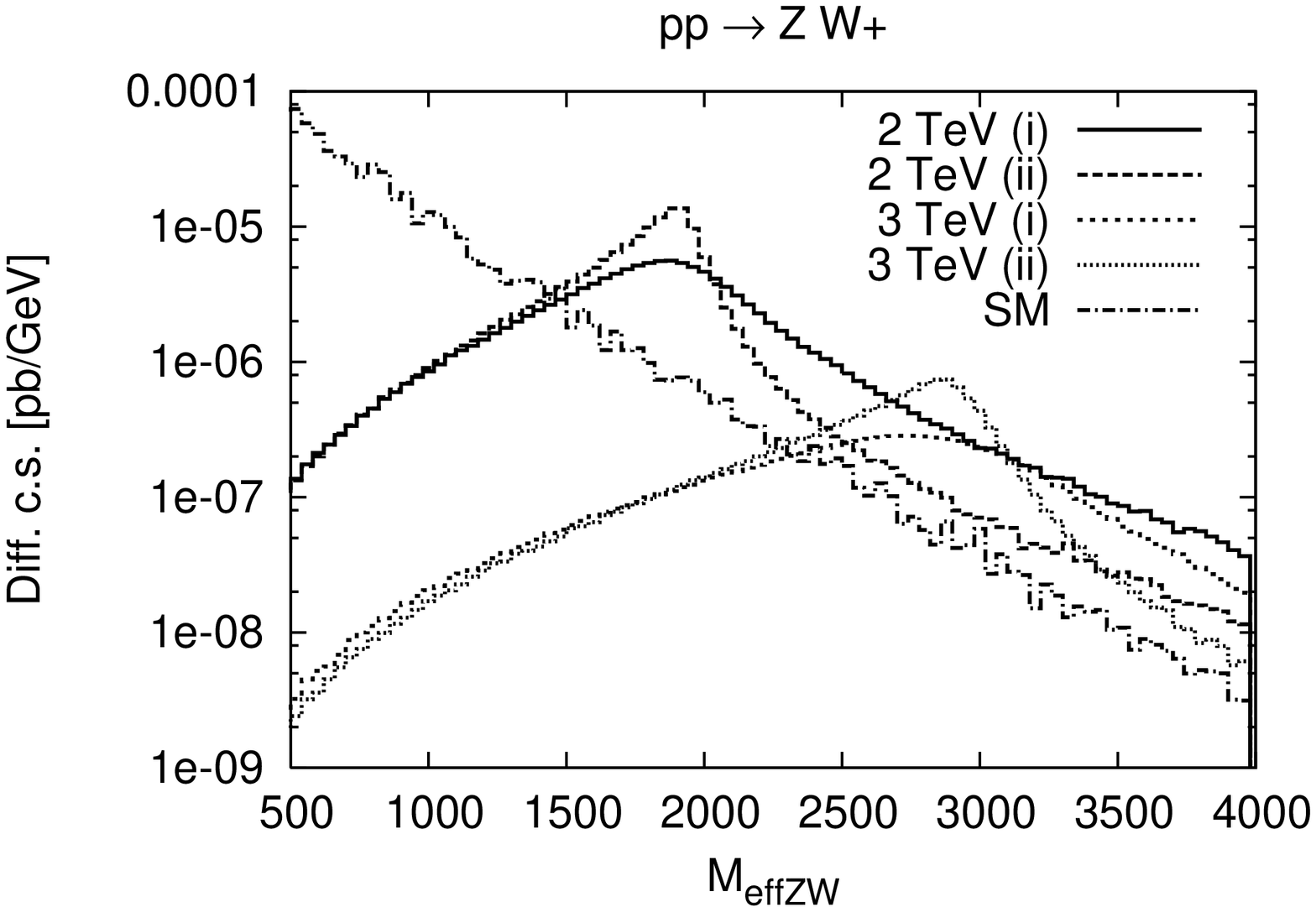}}
\scalebox{0.43}{\includegraphics[angle=270]{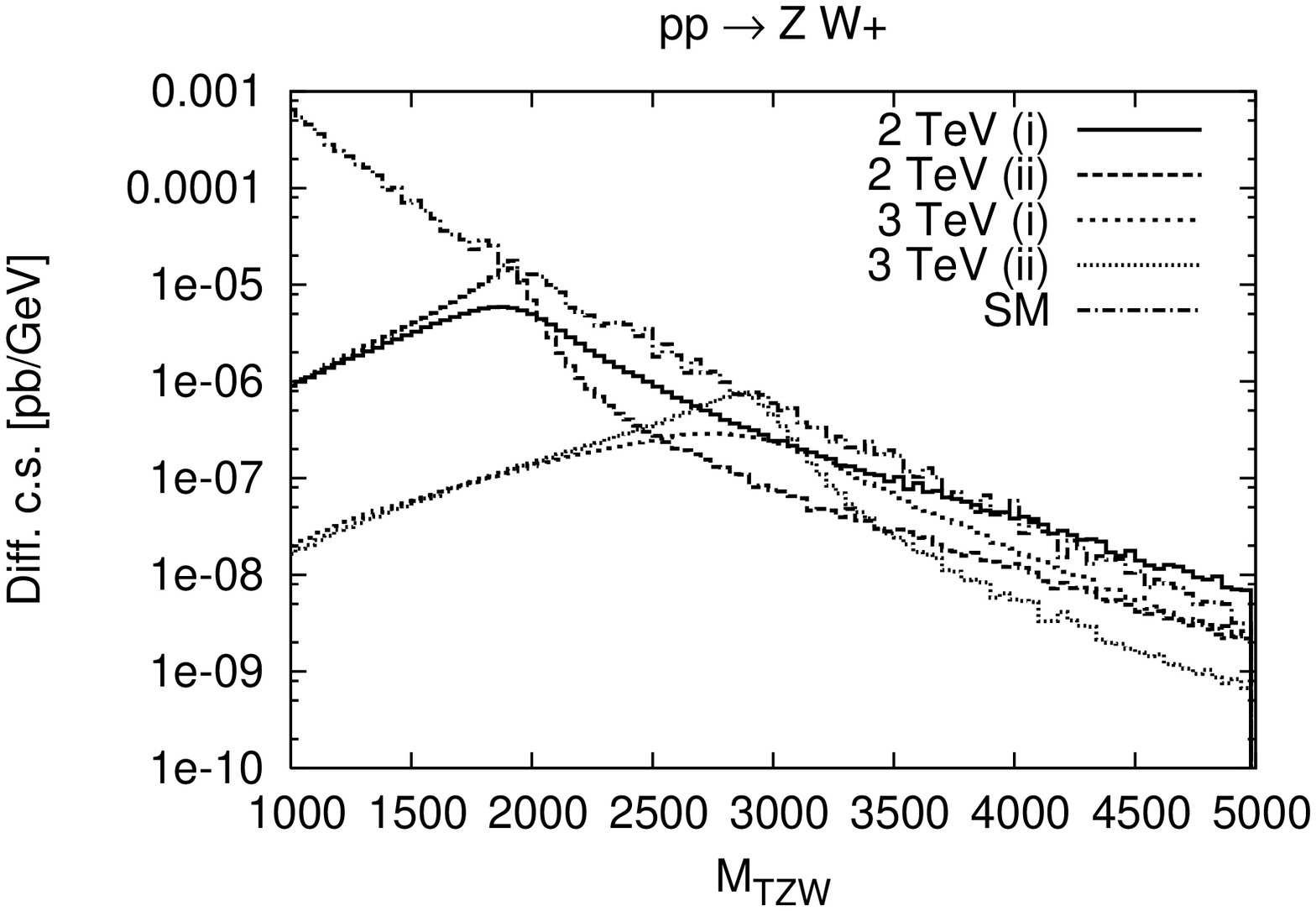}}
\caption{The ${M_{eff}}_{ZW}$ (left) and ${M_T}_{ZW}$ (right) distributions for signal and 
SM background for the process $p p \to Z W^+$ after the cuts: ${p_T}_{Z, W} > 100$~GeV 
and $|y_{Z, W}| < 3$. 
Both case (i) and case (ii) are shown.
\label{MZW.FIG} }
\end{center}
\end{figure}

Since there is at most one missing neutrino in the final state, we can reconstruct the event 
if one makes the assumption $\vec p_\nu \approx \kappa \vec p_e$ 
(see the discussions in the last section). 
On the other hand, it is more straightforward to construct the events 
in the transverse plane. For illustration, we form the following kinematic variables:
\bea
{M_{eff}}_{ZW} &=& {p_T}_{Z} + {p_T}_W \ ,  \\
{M_T}_{ZW} &=& \sqrt{{p_T}_Z^2 + \mz^2} + \sqrt{{p_T}_W^2 + \mw^2} \ . 
\eea
In Fig.~\ref{MZW.FIG} we show the ${M_{eff}}_{ZW}$ and ${M_T}_{ZW}$ differential distributions
for the process $pp \to Z W^+$ for the $\wpri$ signal for cases (i) and (ii), 
and the irreducible SM $WZ$ background.

\subsubsection{Fully leptonic channel}
In the fully leptonic final state we consider the process $pp\to \wpri \to Z W$ followed
by $Z \to \ell \ell$ and $W\to \ell \nu$. 
We take into account the SM $Z W$ going into the same final state as the main source of 
(irreducible) background.  
We select events with the basic cuts
\beq
{p_T}_\ell > 50~{\rm GeV}; \quad {p_T}_{miss} > 50~{\rm GeV}; \quad |\eta_\ell| < 3.
\eeq
In addition to the basic cuts, we apply the following cuts sequentially in 
order to optimally improve signal observation from the background
\begin{itemize}
\item[] $M_{eff}$ cut: $M_{eff} > 1$~TeV (for $\mwpri = 2$~TeV) and $M_{eff} > 1.25$~TeV (for $\mwpri = 3$~TeV).
\item[] $M_T$ cut: $1.5 < {M_T}_{ZW} < 2.5$~TeV (for $\mwpri = 2$~TeV) and $2.4 < {M_T}_{ZW} < 3.6$~TeV 
(for $\mwpri = 3$~TeV).
\end{itemize}
In Table~\ref{pp2llln.TAB} we show the cross-sections (in fb) for the fully leptonic 
signal and background for cases (i) and case (ii) with the above cuts applied. 
\begin{table}
\begin{center}
\caption{
The cross-sections (in fb) for the signal process 
$pp\to \wpri \to Z W  \to \ell \ell \ell \nu $ for case (i) and case (ii),
and SM background, with the cuts applied successively. 
We show cross-sections for $M_\wpri = 2$ and $3$~TeV, and the number of events and significance 
with the luminosity ${\cal L}$ (in ${\rm fb}^{-1}$) as shown for each case. 
\label{pp2llln.TAB}}
\vskip 0.2cm
\begin{tabular}{|c|c|c|c||c|c|c|c|}
\hline 
$2$ TeV&
Basic&
$M_{eff}$&
$M_{T}$&
${\cal L}$&
\# Evts&
$S/B$&
CL\tabularnewline
\hline
\hline 
Case (i)&
$0.13$&
$0.13$&
$0.1$&
$100$&
$10$&
$5$&
$0.9995$\tabularnewline
\hline 
Case (ii)&
$0.17$&
$0.16$&
$0.13$&
$100$&
$13$&
$6.5$&
$>0.9995$\tabularnewline
\hline 
SM $Z\, W$&
$42$&
$0.16$&
$0.02$&
&
$2$&
&
\tabularnewline
\hline
\end{tabular}

\medskip{}

\begin{tabular}{|c|c|c|c||c|c|c|c|}
\hline 
$3$ TeV&
Basic&
$M_{eff}$&
$M_{T}$&
${\cal L}$&
\# Evts&
$S/B$&
CL\tabularnewline
\hline
\hline 
Case (i)&
$0.01$&
$0.01$&
$0.006$&
$1000$&
$6$&
$6$&
$0.99$\tabularnewline
\hline 
Case (ii)&
$0.014$&
$0.01$&
$0.01$&
$1000$&
$10$&
$10$&
$>0.9995$\tabularnewline
\hline 
SM $Z\, W$&
$42$&
$0.05$&
$0.001$&
&
$1$&
&
\tabularnewline
\hline
\end{tabular}
\end{center}
\end{table}
Given the small BR into this final state, it is not surprising that we will need a large 
luminosity to see this signal. The fully leptonic mode is experimentally clean. 
We find that we need ${\cal L}=100~{\rm fb}^{-1}$ 
(${\cal L}=1000~{\rm fb}^{-1}$) for a $2~$TeV ($3~$TeV) $\wpri$
to reach a statistically significant signal (in Poisson statistics).
We turn next to the semi-leptonic mode which has a larger BR and therefore a larger rate.

\subsubsection{Semi-leptonic channel}
We consider below the two semi-leptonic modes $Z\to \ell\ell$, $W\to jj$ and 
$Z\to jj$, $W\to\ell\nu$. 
As explained in detail in Ref.~\cite{Agashe:2007ki}, the two jets may merge into one fat
jet due to the large boost of the parent gauge boson, 
picking up a 1-jet background (in addition to the already mentioned
SM $ZW$ background). 
We now consider the signal identification separately.

\noindent
\underline{$Z\to \ell\ell$, $W\to jj$}: Since there is no missing energy in the event
we can reconstruct the event fully and form the full invariant mass ($M_{inv}$, not just $M_T$).
In addition to the SM $ZW$ background, due to jet merging, we have to contend with $Z\, +$~1-jet
as a source of background. We apply the following cuts to maximize the signal significance:
\begin{itemize}
\item[] Basic cuts: ${p_T}_\ell > 250$~GeV; ${p_T}_j > 500$~GeV; $|\eta_\ell| < 2$; $|\eta_j| < 2$. 
\item[] $M_{eff}$ cut: $M_{eff} > 1$~TeV (for $\mwpri = 2$~TeV) and $M_{eff} > 1.25$~TeV (for $\mwpri = 3$~TeV).
\item[] $M_{inv}$ cut: $1.85 < {M}_{ZW} < 2.15$~TeV (for $\mwpri = 2$~TeV) and 
$2.8 < {M_T}_{ZW} < 3.2$~TeV (for $\mwpri = 3$~TeV).
\item[] Jet-mass cut: $75 < M_{j} < 125$~GeV.
\end{itemize}
In Table~\ref{pp2lljj.TAB} we show the cross-sections as we apply the above cuts successively. 
\begin{table}
\begin{center}
\caption{
The cross-sections (in fb) for the signal process 
$pp\to \wpri \to Z W  \to \ell \ell j j $ for case (i) and case (ii),
and SM background, with the cuts applied successively. 
We show cross-sections for $M_\wpri = 2$ and $3$~TeV, and the number of events and significance 
with the luminosity ${\cal L}$ (in ${\rm fb}^{-1}$) as shown for each case. 
\label{pp2lljj.TAB}}
\vskip 0.2cm
\begin{tabular}{|c|c|c|c|c||c|c|c|c|}
\hline 
2 TeV&
Basic&
$M_{eff}$&
$M_{inv}$&
$M_{j}$&
${\cal L}$&
\# Evts&
$S/B$&
$S/\sqrt{B}$\tabularnewline
\hline
\hline 
Case (i)&
$0.4$&
$0.4$&
$0.16$&
$0.13$&
$1000$&
$130$&
$0.2$&
$5$\tabularnewline
\hline 
Case (ii)&
$0.5$&
$0.48$&
$0.38$&
$0.3$&
$300$&
$90$&
$0.5$&
$6.4$\tabularnewline
\hline 
SM $ZW$&
$130$&
$0.5$&
$0.05$&
$0.04$&
&
$40$, $12$&
&
\tabularnewline
\hline 
SM $Z$ + 1j&
$3600$&
$63$&
$2.1$&
$0.63$&
&
$630$, $190$&
&
\tabularnewline
\hline
\end{tabular}

\medskip{}

\begin{tabular}{|c|c|c|c|c||c|c|c|c|}
\hline 
3 TeV&
Basic&
$M_{eff}$&
$M_{inv}$&
$M_{j}$&
${\cal L}$&
\# Evts&
$S/B$&
$S/\sqrt{B}$\tabularnewline
\hline
\hline 
Case (i)&
$0.03$&
$0.03$&
$0.01$&
$-$&
$1000$&
$10$&
$0.07$&
$0.8$\tabularnewline
\hline 
Case (ii)&
$0.04$&
$0.04$&
$0.03$&
$-$&
$1000$&
$30$&
$0.22$&
$2.6$\tabularnewline
\hline 
SM $ZW$&
$130$&
$0.16$&
$0.006$&
$-$&
&
$6$&
&
\tabularnewline
\hline 
SM $Z$ + 1j&
$3600$&
$25$&
$0.13$&
$-$&
&
$130$&
&
\tabularnewline
\hline
\end{tabular}
\end{center}
\end{table}

\noindent
\underline{$Z\to jj$, $W\to \ell\nu$}:
In addition to the SM $ZW$ background, due to jet merging, we have to contend with $W\, +$~1-jet
as a source of background. 
We apply the following cuts to maximize significance:
\begin{itemize}
\item[] Basic cuts: ${p_T}_\ell > 50$~GeV; $\etmiss > 50$~GeV; $|\eta_\ell| < 1$; $|\eta_j| < 1$. 
\item[] $M_{eff}$ cut: $M_{eff} > 1$~TeV (for $\mwpri = 2$~TeV) and $M_{eff} > 1.25$~TeV (for $\mwpri = 3$~TeV).
\item[] $M_{T}$ cut: $1.8 < {M_T}_{ZW} < 2.2$~TeV (for $\mwpri = 2$~TeV) and 
$2.8 < {M_T}_{ZW} < 3.2$~TeV 
(for $\mwpri = 3$~TeV).
\item[] Jet-mass cut: $75 < M_{j} < 125$~GeV.
\end{itemize}
In Table~\ref{pp2jjln.TAB} we show the cross-sections as we apply the above cuts successively. 
\begin{table}
\begin{center}
\caption{
The cross-sections (in fb) for the signal process 
$pp\to \wpri \to Z W  \to j j \ell \etmiss $ for case (i) and case (ii),
and SM background, with the cuts applied successively. 
We show cross-sections for $M_\wpri = 2$ and $3$~TeV, and the number of events and significance 
with the luminosity ${\cal L}$ (in ${\rm fb}^{-1}$) as shown for each case. 
\label{pp2jjln.TAB}}
\vskip 0.2cm
\begin{tabular}{|c|c|c|c|c||c|c|c|c|}
\hline 
2 TeV&
Basic&
$M_{eff}$&
$M_{T}$&
$M_{jet}$&
${\cal L}$&
\# Evts&
$S/B$&
$S/\sqrt{B}$\tabularnewline
\hline
\hline 
Case (i)&
$1$&
$1$&
$0.38$&
$0.3$&
$1000$&
$300$&
$0.1$&
$5.3$\tabularnewline
\hline 
Case (ii)&
$1.3$&
$1.2$&
$0.64$&
$0.5$&
$300$&
$150$&
$0.16$&
$4.9$\tabularnewline
\hline 
SM $ZW$&
$320$&
$1.2$&
$0.04$&
$0.03$&
&
$30$, $9$&
&
\tabularnewline
\hline 
SM $W$ + 1j&
$3.1\times10^{4}$&
$220$&
$10.5$&
$3.2$&
&
$3200$, $950$&
&
\tabularnewline
\hline
\end{tabular}

\medskip{}

\begin{tabular}{|c|c|c|c|c||c|c|c|c|}
\hline 
3 TeV&
Basic&
$M_{eff}$&
$M_{T}$&
$M_{jet}$&
${\cal L}$&
\# Evts&
$S/B$&
$S/\sqrt{B}$\tabularnewline
\hline
\hline 
Case (i)&
$0.08$&
$0.08$&
$0.016$&
$-$&
$1000$&
$16$&
$0.02$&
$0.6$\tabularnewline
\hline 
Case (ii)&
$0.1$&
$0.1$&
$0.04$&
$-$&
$1000$&
$40$&
$0.06$&
$1.5$\tabularnewline
\hline 
SM $ZW$&
$320$&
$0.4$&
$0.002$&
$-$&
&
$2$&
&
\tabularnewline
\hline 
SM $W$ + 1j&
$3.1\times10^{4}$&
$89$&
$0.68$&
$-$&
&
$680$&
&
\tabularnewline
\hline
\end{tabular}
\end{center}
\end{table}

In the semi-leptonic channels presented above, 
we find for a $2~$TeV $\wpri$ that we need a luminosity of
${\cal L}=1000~{\rm fb}^{-1}$ and ${\cal L}=300~{\rm fb}^{-1}$ for cases (i) and (ii) 
respectively.
We thus see that the $tb$ final state discussed earlier
which requires about $100~{\rm fb}^{-1}$,
offers a more promising 
channel for the $W'$ signal observation for case (i) 
as compared to the semileptonic W/Z channels. 
For a $3~$TeV $\wpri$ we find that 
the QCD background is substantial and
the signal-to-background ratio is at a level of a few percent, rendering the
signal observation unlikely. 
Techniques to beat down this reducible QCD background can be beneficial here.

The semi-leptonically decaying neutral electroweak KK gauge boson ($\zpri$) 
also decays into the $jj\ell\nu$ final state and its detectability has already been  
discussed in Ref.~\cite{Agashe:2007ki}. 

\subsection{$W h$ final state}
\begin{figure}
\begin{center}
\scalebox{0.43}{\includegraphics[angle=270]{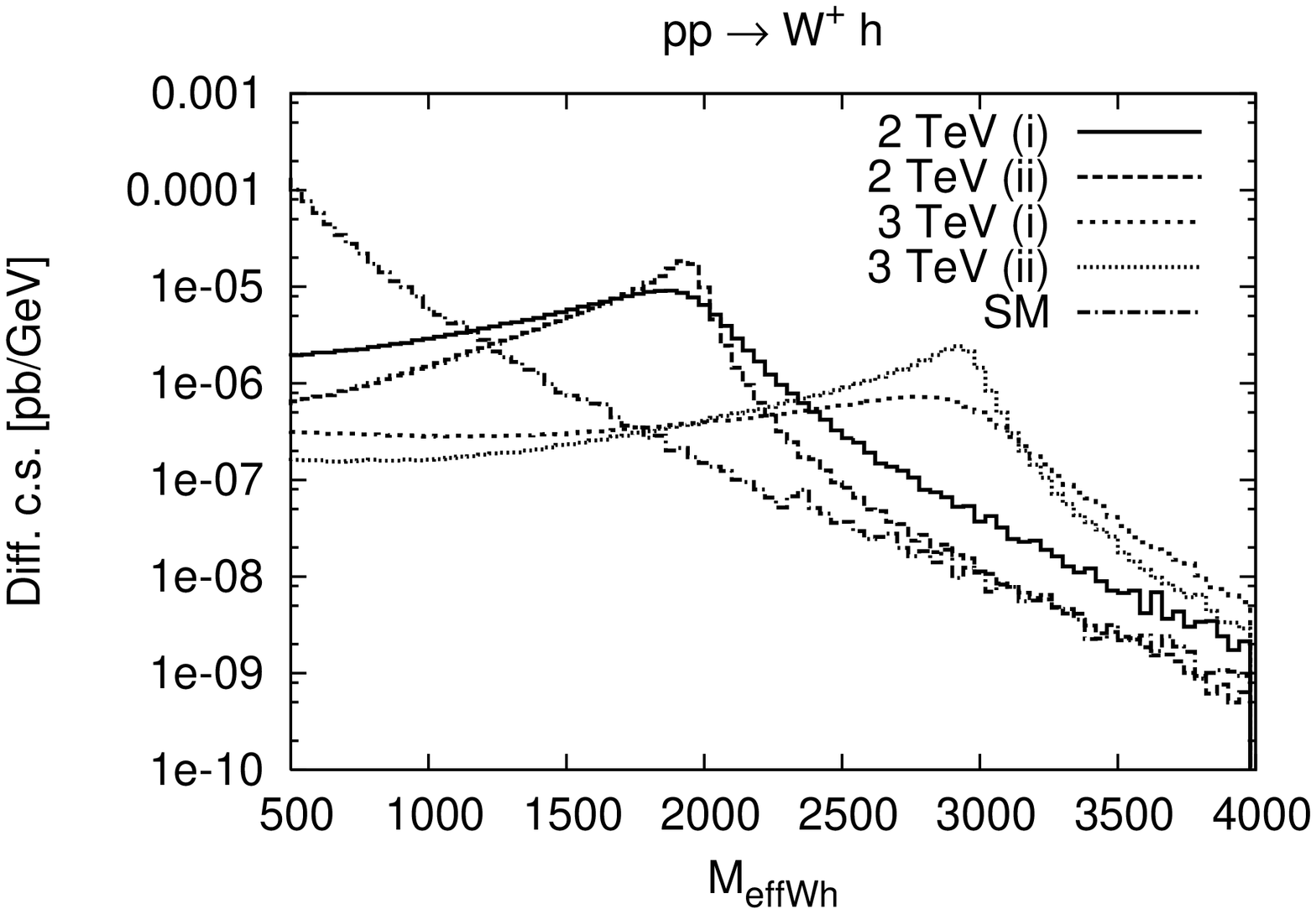}}
\scalebox{0.43}{\includegraphics[angle=270]{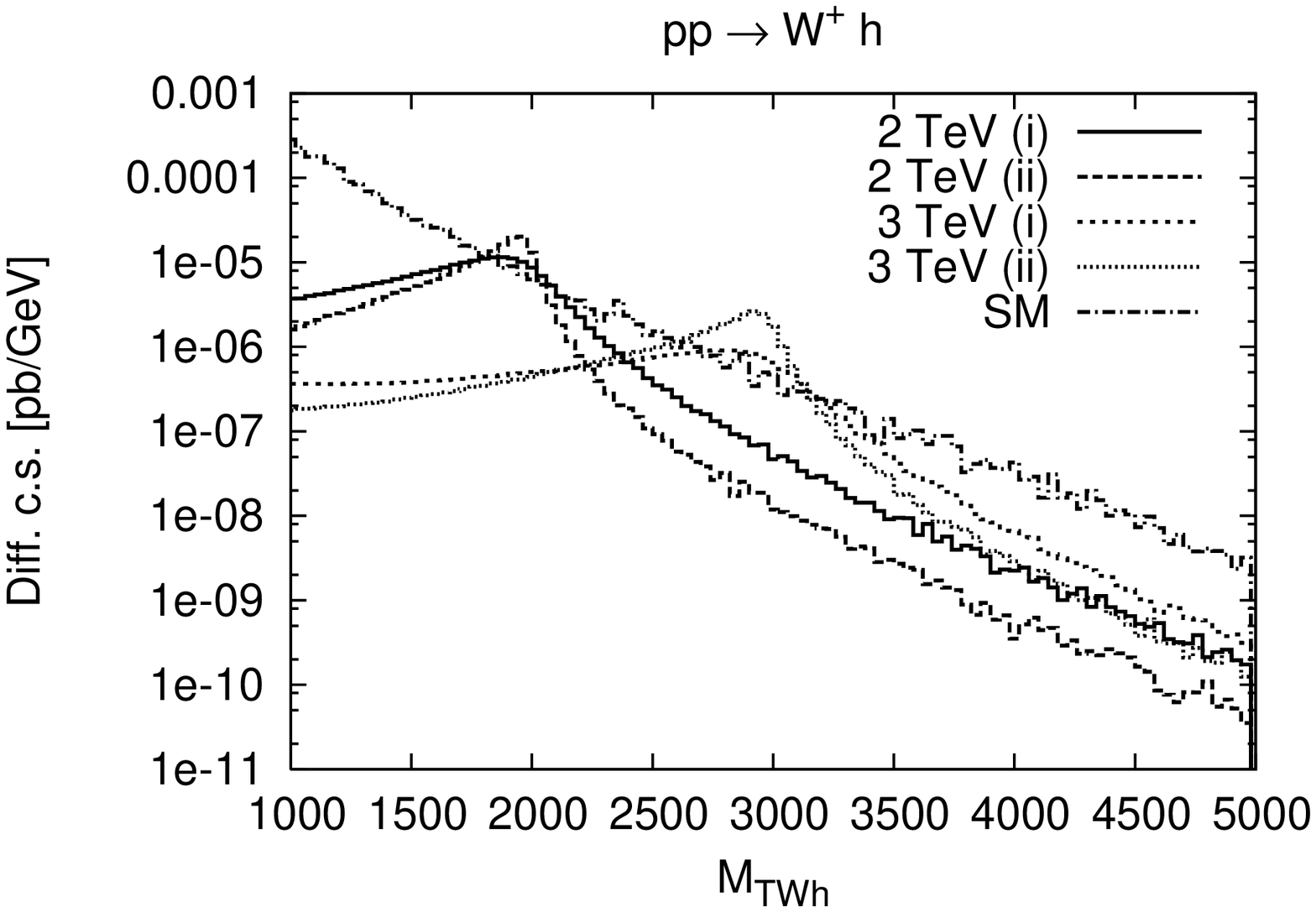}}
\caption{The ${M_{eff}}_{Wh}$ (left) and ${M_T}_{Wh}$ (right) distributions for signal and 
SM background for the process $p p \to W^+ h$ after the cuts: ${p_T}_{W, h} > 100$~GeV 
and $|y_{W, h}| < 3$. 
Both case (i) and case (ii) are shown.
\label{MWh.FIG} }
\end{center}
\end{figure}

Similar to the study in the last section, we again first  
form the following kinematic variables in order to help separate the signal from background:
\bea
{M_{eff}}_{Wh} &=& {p_T}_{W} + {p_T}_h \ ,  \\
{M_T}_{Wh} &=& \sqrt{{p_T}_W^2 + \mw^2} + \sqrt{{p_T}_h^2 + \mh^2} \ . 
\eea
In Fig.~\ref{MWh.FIG} we show the ${M_{eff}}_{Wh}$ and ${M_T}_{Wh}$ differential distributions
for the process $pp \to W^+ h$ for the $\wpri$ signal for cases (i) and (ii), 
and the irreducible SM $Wh$ background.
We see that the signal stands comfortably over the background, and with suitably 
chosen cuts we expect to obtain a good significance.
Since $\mh$ is unknown, we will take two representative cases: $\mh = 120$~GeV and
$150$~GeV. In the former case the dominant decay mode of the $h$ 
will be to $b \bar b$, while in the latter, to $W^+ W^-$. 
We will consider each of these cases in turn. 

We would like to note that we do not
consider the studies here as the Higgs boson discovery channels. Instead, we should
consider them as case studies for illustration since we should have had the knowledge
about the Higgs properties when our proposed searches are undertaken at the LHC.

\subsubsection{$\mh = 120$~GeV: $h\to b\bar b$, $W\to \ell\nu$}
For this case with a relatively low mass, 
we estimate that $BR(h\to b\bar b) \approx 0.7$. 
Due to collimation of the decay products of the Higgs,
the two b-jets could merge, and we therefore pick-up $W + 1$ jet as a source of background.
We apply the following cuts to maximize significance:
\begin{itemize}
\item[] Basic cuts: ${p_T}_\ell > 50$~GeV; $\etmiss > 50$~GeV; ${p_T}_{(bb)} > 100$~GeV; $|\eta_\ell| < 1$; $|\eta_j| < 1$. 
\item[] $M_{eff}$ cut: $M_{eff} > 1$~TeV (for $\mwpri = 2$~TeV) and $M_{eff} > 1.25$~TeV (for $\mwpri = 3$~TeV).
\item[] $M_{T}$ cut: $1.8 < {M_T}_{Wh} < 2.2$~TeV (for $\mwpri = 2$~TeV) and $2.8 < {M_T}_{Wh} < 3.2$~TeV 
(for $\mwpri = 3$~TeV).
\item[] b-tag: Due to collimation, we may not be able to resolve the two $b$-jets, 
and we therefore demand 
only one b-tag. The efficiency for at least one tagged $b$ is $\epsilon_b * (2-\epsilon_b)$. Here, 
we take the light jet rejection ratio $R_j = 20$, which, as noted earlier, 
will likely be improved upon.
\end{itemize}
In addition to the above cuts, we could apply a jet-mass cut on the collimated $b$-jet system which 
will peak around $\mh$, and can be used to distinguish it from a light-jet. 
Doing so would improve the significance over that shown here.

In Table~\ref{pp2Wh2lnbb.TAB} we show the cross-sections as we apply the above cuts successively. 
\begin{table}
\begin{center}
\caption{
The cross-sections (in fb) for the signal process 
$pp\to \wpri \to W h  \to \ell \etmiss b \bar b$ for case (i) and case (ii),
and SM background, with the cuts applied successively. 
We show cross-sections for $M_\wpri = 2$ and $3$~TeV, and the number of events and significance 
with the luminosity ${\cal L}$ (in ${\rm fb}^{-1}$) as shown for each case. 
\label{pp2Wh2lnbb.TAB}}
\begin{tabular}{|c|c|c|c|c||c|c|c|c|}
\hline 
2 TeV&
Basic&
$M_{eff}$&
$M_{T}$&
b-tag&
${\cal L}$&
\# Evts&
$S/B$&
$S/\sqrt{B}$\tabularnewline
\hline
\hline 
Case (i)&
$1.8$&
$1.5$&
$0.55$&
$0.35$&
$100$&
$35$&
$0.65$&
$4.8$\tabularnewline
\hline 
Case (ii)&
$1.6$&
$1.5$&
$0.8$&
$0.5$&
$100$&
$50$&
$1$&
$6.4$\tabularnewline
\hline 
SM $Wh$&
$43$&
$0.35$&
$0.016$&
$0.01$&
&
$1$&
&
\tabularnewline
\hline 
SM $W$ + 1j&
$3.1\times10^{4}$&
$220$&
$10.5$&
$0.53$&
&
$53$&
&
\tabularnewline
\hline
\end{tabular}

\medskip{}

\begin{tabular}{|c|c|c|c|c||c|c|c|c|}
\hline 
3 TeV&
Basic&
$M_{eff}$&
$M_{T}$&
b-tag&
${\cal L}$&
\# Evts&
$S/B$&
$CL$\tabularnewline
\hline
\hline 
Case (i)&
$0.26$&
$0.19$&
$0.04$&
$0.03$&
$300$&
$9$&
$1$&
$0.99$\tabularnewline
\hline 
Case (ii)&
$0.33$&
$0.3$&
$0.12$&
$0.08$&
$300$&
$24$&
$2.4$&
$>0.9995$\tabularnewline
\hline 
SM $Wh$&
$43$&
$0.13$&
$0.001$&
$6\times10^{-4}$&
&
$0.2$&
&
\tabularnewline
\hline 
SM $W$ + 1j&
$3.1\times10^{4}$&
$89$&
$0.68$&
$0.03$&
&
$9$&
&
\tabularnewline
\hline
\end{tabular}
\end{center}
\end{table}
As expected we find a better significance for case (ii) since the BR is larger. 
We will need ${\cal L}=100~{\rm fb}^{-1}$ (${\cal L}=300~{\rm fb}^{-1}$) 
for a $2~$TeV ($3~$TeV) $\wpri$ to reach a good statistical significance.
Improving the $b$-tagging performance (by achieving larger $R_j$) will help reduce 
the $W$~+~1~jet background and will better the significance.

\subsubsection{$\mh = 150$~GeV: $h\to WW\to \ell\nu j j$, $W\to j j$}
For this case with a higher mass, we estimate that $BR(h\to WW) \approx 0.7$. 
Due to collimation of the jets from the $W$, we will not demand that separate jets be 
reconstructed, but rather treat it as a single jet. We will refer to the merged
jet closer to the leptonic $W$ as the near-jet $j_N$ and the merged jet on the other side
as the far-jet $j_F$.

We require that there be a jet close to the lepton with $\Delta_{\ell j_N} < 0.9$,
and require ${M_T}_{W j_N}$ to be around $M_h$, which will 
reduce the $W+2j$ background.
In addition to the irreducible SM $W h$ background, SM $h+1j$ also remains as a background, 
which we will include. 
In order to differentiate between a light-jet and the $W$-jet we will apply the jet-mass cut 
as explained in Ref.~\cite{Agashe:2007ki}.

We apply the following cuts to maximize significance:
\begin{itemize}
\item[] Basic cuts: ${p_T}_\ell > 25$~GeV; $\etmiss > 25$~GeV; ${p_T}_{j_N} > 50$~GeV; ${p_T}_{j_F} > 100$~GeV; $|\eta_\ell| < 3$; $|\eta_{j_N}| < 3$; $|\eta_{j_F}| < 3$. 
\item[] $M_{eff}$ cut: $M_{eff} > 1$~TeV (for $\mwpri = 2$~TeV) and $M_{eff} > 1.25$~TeV (for $\mwpri = 3$~TeV).
\item[] $M_{T}$ cuts: $100 < {M_T}_{W j_N} < 190$~GeV (around $\mh$);  $1.8 < {M_T}_{W j_N j_F} < 2.2$~TeV (for $\mwpri = 2$~TeV) and $2.8 < {M_T}_{W j_N j_F} < 3.2$~TeV 
(for $\mwpri = 3$~TeV).
\item[] Jet-mass cut: $75 < M_j < 125$~GeV, on both $j_N$ and $j_F$ for $\mwpri = 2$~TeV, with 
an acceptance of $0.87$ for the $W$-jet and $0.3$ for a light-jet~\cite{Agashe:2007ki}. 
For $\mwpri = 3$~TeV, we apply the jet-mass cut only 
on $j_N$ since its performance in $j_F$ might deteriorate owing to increased collimation. 
\end{itemize}

In Table~\ref{pp2Wjjjj.TAB} we show the cross-sections as we apply the above cuts successively. 
\begin{table}
\begin{center}
\caption{
The cross-sections (in fb) for the signal process 
$pp\to \wpri \to W h  \to (jj) WW \to (jj) \ell \etmiss (jj)$ for case (i) and case (ii),
and SM background, with the cuts applied successively. 
We show cross-sections for $M_\wpri = 2$ and $3$~TeV, and the number of events and significance 
with the luminosity ${\cal L}$ (in ${\rm fb}^{-1}$) as shown for each case.
\label{pp2Wjjjj.TAB}}
\begin{tabular}{|c|c|c|c|c||c|c|c|c|}
\hline 
2 TeV&
Basic&
$M_{eff}$&
$M_{T}$&
$M_{jet}$&
${\cal L}$&
\# Evts&
$S/B$&
CL\tabularnewline
\hline
\hline 
Case (i)&
$1.6$&
$1.3$&
$0.43$&
$0.34$&
$100$&
$34$&
$4$&
$\gg0.9995$\tabularnewline
\hline 
Case (ii)&
$2.1$&
$1.9$&
$0.9$&
$0.7$&
$100$&
$70$&
$7$&
$\gg0.9995$ \tabularnewline
\hline 
SM $W\, h$&
$26$&
$0.31$&
$0.014$&
$0.01$&
&
$1$&
&
\tabularnewline
\hline 
SM $h+1j$&
$220$&
$2$&
$0.07$&
$0.02$&
&
$2$&
&
\tabularnewline
\hline 
SM $W+2j$&
$3\times10^{4}$&
$36$&
$0.62$&
$0.06$&
&
$6$&
&
\tabularnewline
\hline
\end{tabular}

\medskip{}

\begin{tabular}{|c|c|c|c|c||c|c|c|c|}
\hline 
3 TeV&
Basic&
$M_{eff}$&
$M_{T}$&
$M_{jet}$&
${\cal L}$&
\# Evts&
$S/B$&
CL\tabularnewline
\hline
\hline 
Case (i)&
$0.22$&
$0.17$&
$0.04$&
$0.035$&
$300$&
$11$&
$2$&
$0.9987$\tabularnewline
\hline 
Case (ii)&
$0.3$&
$0.26$&
$0.1$&
$0.09$&
$300$&
$27$&
$4$&
$\gg0.9995$\tabularnewline
\hline 
SM $W\, h$&
$26$&
$0.12$&
$8\times10^{-4}$&
$7\times10^{-4}$&
&
$0.2$&
&
\tabularnewline
\hline 
SM $h+1j$&
$220$&
$0.72$&
$5\times10^{-3}$&
$2\times10^{-3}$&
&
$0.6$&
&
\tabularnewline
\hline 
SM $W+2j$&
$3\times10^{4}$&
$4.1$&
$0.05$&
$0.015$&
&
$4.5$&
&
\tabularnewline
\hline
\end{tabular}
\end{center}
\end{table}
Similar to the previous case, we find that we will need 
${\cal L}=100~{\rm fb}^{-1}$ (${\cal L}=300~{\rm fb}^{-1}$) 
for a $2~$TeV ($3~$TeV) $\wpri$ to reach a good statistical significance.
The $S/B$ is found to be quite adequate for a signal discovery, 
and the reach is limited by signal statistics. 

\subsection{$W'\to \ell \, \nu$ final state}
In spite of the unique signal kinematics, we
expect the signal event rate to be quite small for this final state given the 
tiny branching ratio for this mode. 
Nevertheless, for completeness, we show in Table~\ref{pp2ln.TAB}
for $\mwpri = 2$~TeV the cross-sections for this mode 
after the following cuts:
\begin{itemize}
\item[] Basic cuts: ${p_T}_\ell > 100$~GeV; $\etmiss > 100$~GeV; $|\eta_\ell| < 3$.
\item[] $M_{eff}$ cut: $M_{eff} > 1$~TeV.
\item[] $M_{T}$ cut: $1.5 < {M_T}_{\ell \nu} < 2.5$~TeV.
\end{itemize}
%
We include the SM $W^\pm$ exchange irreducible background. 
\begin{table}
\begin{center}
\caption{
The cross-sections (in fb) for the signal process 
$pp\to \wpri \to \ell \nu$ for case (i) and case (ii),
and SM background, with the cuts applied successively,  
for $M_\wpri = 2$~TeV. 
\label{pp2ln.TAB}}
\vskip 0.2cm
\begin{tabular}{|c|c|c|c|}
\hline 
2 TeV&
Basic&
$M_{eff}$&
$M_{T}$\tabularnewline
\hline
\hline 
Case (i)&
$0.04$&
$0.024$&
$0.012$\tabularnewline
\hline 
Case (ii)&
$0.05$&
$0.04$&
$0.02$\tabularnewline
\hline 
SM $W$&
$4\times10^{3}$&
$6.9$&
$0.44$\tabularnewline
\hline
\end{tabular}
\end{center}
\end{table}
As expected, the signal rate is rather low in comparison with the irreducible SM background.
We thus do not expect this mode to be detectable. The modes explored 
in the previous subsections have much better reach as we have demonstrated.

\section{Comparison to Technicolor Studies}
\label{TC}

Based on the AdS/CFT correspondence,
the warped extra dimensional model is 
conjectured to be dual to purely $4D$ strong dynamics
being involved in EWSB, such
as technicolor or composite Higgs models.
So, we expect similar
signals for the two scenarios and therefore
it is useful to compare
the extensive technicolor studies in the literature (see
Ref.~\cite{Hill:2002ap}
for a review)
with our current work on signals for electroweak KK gauge
bosons 
in warped extra dimension (including neutral
case studied earlier in Ref.~\cite{Agashe:2007ki}).

We begin with the details of this duality which
will enable us to compare the signals
that we studied to technicolor studies. The $5D$ model
corresponds to a  $4D$ theory with {\em two} sectors\footnote{See
reference \cite{Contino:2006nn}
for a two-site description 
%
%
of the $5D$ model (including the couplings to the
heavy new particles) along these lines.}. There
is a sector which is strongly coupled, with the strength
of the couplings in this sector
remaining approximately constant 
over the Planck-weak hierarchy, i.e., it is a quasi-conformal
theory. Conformal invariance is broken
at the TeV
scale, resulting in a tower
of composite
(bound) states starting at $\sim$ TeV scale. 
The 2nd sector consists of particles
external to this conformal sector or elementary
(as opposed to the composites of the strong sector above). However, 
these 2 sectors are not isolated, i.e.,
they do couple to each other. As a result,
the elementary particles (external
to the CFT sector) mix with the CFT composites and the
mass eigenstates
(physical states) are admixtures of the two
sets of particles. These physical
states correspond to the zero and KK modes of the $5D$ theory.

Furthermore, 
the location of a mode in the extra dimension is dual to
the amount or degree of compositeness (in the sense
of the elementary-composite mixture
above) of the corresponding state
in the $4D$ theory. Specifically, modes which are localized near
the Planck (TeV) brane are interpreted as states which are
mostly elementary (composite). Thus, the light SM fermions are
mostly elementary, whereas the top quark, Higgs 
(including unphysical Higgs or
longitudinal $W/Z$) and all
KK's are mostly composites (the SM gauge bosons with
a flat profile are in-between in terms of compositeness).
Roughly speaking, the KK tower 
of the $5D$ theory then corresponds to the tower
of (massive) composites (``hadrons'') in the $4D$ theory.
As discussed earlier, the coupling of a set of modes
of the $5D$ theory is proportional to the
overlap of the corresponding profiles
in the extra dimension, i.e., it is large 
if all the modes of this set are localized near the
TeV (or Planck) brane and small if some modes
are localized near the Planck brane while others
are localized near TeV brane. In the dual $4D$ theory,
the 1st situation correspond to all the particles
of the set being mostly composite (or mostly
elementary), clearly resulting
in a large coupling between these particles, while the 2nd
situation involves some particles
which are mostly elementary and others which are mostly composites
(thus
accounting for the small coupling).

We now compare the nature of couplings
and hence the decay channels in the warped extra dimensional
model that we studied to the case of technicolor theories studied previously.
First of all, the decays to physical Higgs bosons ($+ W/Z$)
for the electroweak KK's that we studied are new compared to technicolor
studies.
The reason is that in technicolor theories (at least in the minimal
models), there would not
be a light Higgs since the idea of technicolor models
is that the strong dynamics
directly or spontaneously breaks EW symmetry.
Equivalently, $WW$ scattering is 
unitarized
by exchange of spin-1 bound states (techni-$\rho$'s)
instead of by a (light) Higgs.
On the other hand the warped extra dimensional model that we studied
(with a light
Higgs in the spectrum) is dual to composite Higgs models
in $4D$, i.e., where strong
dynamics does not directly break EW symmetry.
Rather, the strong dynamics
produces a light composite Higgs which then acquires
a vev in the low energy theory to break EW symmetry.

However, the decays
of electroweak KK's to $WZ$ or $WW$ 
and production of KK's in vector boson fusion
are
(qualitatively) similar to 
those studied 
in the 
technicolor literature for the following reason.
Recall that the decays of electroweak KK's
to $WZ$ and $WW$ are dominated by {\em longitudinal}
polarizations
of the latter, that too with couplings  
which are
enhanced relative to the SM.
Since longitudinal $W/Z$ are equivalent to
unphysical
Higgs, it is clear (based on the above discussion) that this
coupling is dual to a self-coupling
of three composites (techni-$\rho$
with composite Goldstones) in the $4D$ theory
and thus is expected to be large. Clearly,
such a coupling is a general characteristic
of EWSB originating
from strong dynamics and is present in
all technicolor models studied in the literature.
Of course, the details of these couplings at the
quantitative level will be different in the
$5D$ model than in the technicolor case (see
Ref.~\cite{Giudice:2007fh} for a model-independent parametrization of couplings
in composite Higgs models).

For the case of couplings of gauge KK's to fermions, it is convenient
to consider two pieces or contributions 
in the formula for this coupling (see Table~\ref{ovlap_ffG.TAB}) as follows. 
It can be shown that the piece
$\propto 1 / \xi$ comes from overlap of profiles near the Planck or UV brane.
This part of the coupling is dual to the
SM fermion first coupling to the photon/$W$/$Z$ (external to strong dynamics)
which then mixes with the composite techni-$\rho$
(analog of photon-$\rho$ mixing in 
QCD). Clearly, this piece of the coupling is present in technicolor models 
studied in the literature as well
and is flavor universal.

The contribution to the coupling of SM fermions to gauge KK's
$\propto \xi$ originates from the overlap of profiles near
the TeV or IR brane. In the $4D$ theory,
this part of the coupling corresponds to a
{\em direct} coupling of SM fermions to the techni-$\rho$, i.e., 
a coupling involving the {\em composite} component of the techni-$\rho$
(as opposed to the coupling via techni-$\rho$'s mixing with
external gauge bosons).
Clearly such a contribution
arises from (partial) compositeness of the SM
fermions themselves and is 
of similar size to the fermions' coupling to the Higgs (which is another composite).
Thus, this piece of the coupling is $\propto 4D$ or SM Yukawa coupling
and is therefore flavor-dependent. 

This second
contribution to the SM fermion
coupling to KK's is absent in ``extended technicolor'' (ETC), which
is the mechanism used in traditional technicolor models
to generate fermion masses (instead
of partial
compositeness of SM fermions
as described above). In detail, in ETC, the SM fermion masses
originate from the
coupling of {\em two} SM fermions to
(a scalar operator of) strong dynamics such that
there is no mixing of external fermions with composite fermions,
unlike in the partial compositeness case which involves
coupling of a {\em single} external fermion
to (a fermionic operator of) strong dynamics.
In any case,
this piece of the coupling is irrelevant
for production of gauge KK's via Drell-Yan (DY) process since that
involves (dominantly) light fermions, whereas it is relevant for decays
of gauge KK's into heavier SM fermions (top/bottom). Therefore,
DY production of gauge KK's is (at 
least qualitatively) similar to that of
techni-$\rho$'s in technicolor, whereas decays to
top quarks are different than in the simplest technicolor models
with ETC.

In general,
using
the warped extra dimension framework has
the advantage that we have a concrete,
weakly coupled model so that we can ensure
that we have a consistent set of couplings.
In contrast,
most technicolor studies simply used a parametrization 
for the various couplings rather
than a 
%
%
well-defined model,
although one could conceivably have 
such a 
%
%
model for these couplings by rescaling QCD data,
assuming the strong dynamics is QCD-like.

Other differences between our analyses and
earlier studies of technicolor are as follows.
Most of the technicolor
studies did not go beyond $\sim 2.5$ TeV
mass for the techni-$\rho$'s,
although the heavier end of the mass range was preferred by 
constraints from EWPT
(specifically the $S$ parameter),
while
we have considered signals for
electroweak KK's up to $3$ TeV.
Finally, 
the semileptonic decay of the $WW$ or $WZ$ has 
not been studied in detail in the technicolor
context, especially the use of 
jet mass cut to discriminate a $W/Z$ jet from
a QCD jet.

\section{Discussion and Conclusions}
\label{conclude}

In the past few years, it has been shown that 
the framework of a warped extra dimension with SM fields in the bulk
can address many of the puzzles of nature. 
Thus, this framework
is a very attractive extension of the
SM (perhaps as compelling
as SUSY).  As the LHC has started,
it is very crucial to study in this framework robust signals from the {\em direct} 
production at the LHC of the new particles, namely the KK excitations
of the SM. 
Over the last year or so, such analyses
have been performed for the KK gluon, graviton, $Z$ 
and some fermions. 
Here, we
continue this program with a study of the {\em charged} electroweak
KK gauge bosons ($W'$) thus completing the study of spin-$1$ states in this
framework.
We summarize in Table~\ref{TAB:sum} the LHC reach
for the two $t_R$ 
cases discussed in Sec.~\ref{outline} with 
representations shown in Eq.~(\ref{tRdef.EQ}),
extracting from the detailed analysis we presented in Sec.~\ref{signal}
the best channel for each of the $\wpri$ decay modes. We give
the luminosity and resulting significance for the mass shown.
We find that we can get a sensitivity of $2~(3)$ TeV masses
with an integrated luminosity of about $100~(300)$ fb$^{-1}$, 
which although slightly better is comparable to the
KK $Z$ reach obtained in Ref.~\cite{Agashe:2007ki}. 
%
\begin{table}
\begin{center}
\caption{Summary of the best channel for each of the $\wpri$ decay modes, 
giving the luminosity and significance for the mass shown, 
in the two $t_R$ coupling scenarios of Case (i) and (ii).
For the $t\, b$ channel the numbers without (and with) the reducible $t\bar t$ 
background are shown.
\label{TAB:sum}}
\begin{tabular}{|c|c|c|c|c|}
\hline 
Case (i):  Channel&
$\mwpri$ (TeV)&
${\cal L}$ ($fb^{-1}$)&
${S}/{B}$&
${S}/{\sqrt{B}}$\tabularnewline
\hline
\hline 
$t\, b\to\ell\nu b\bar{b}$&
$3$&
$300$&
$5.8\,(0.9)$&
$0.995\,(0.95)$ CL\tabularnewline
\hline 
$Z\, W\to\ell\ell\ell\nu$&
$3$&
$1000$&
$6$&
$0.99$ CL\tabularnewline
\hline 
$m_{h}=120$: $W\, h\to\ell\nu b\bar{b}$ &
$3$&
$300$&
$1$&
$0.99$ CL\tabularnewline
\hline 
$m_{h}=150$: $W\, h\to(jj)\,\ell\nu\,(jj)$ &
$3$&
$300$&
$2$&
$0.9987$ CL\tabularnewline
\hline
\end{tabular}

\begin{tabular}{|c|c|c|c|c|}
\hline 
Case (ii):  Channel&
$\mwpri$ (TeV)&
${\cal L}$ ($fb^{-1}$)&
${S}/{B}$&
${S}/{\sqrt{B}}$\tabularnewline
\hline
\hline 
$t\, b\to\ell\nu b\bar{b}$&
$2$&
$1000$&
$0.4\,(0.2)$&
$3.4\,(2.5)$ $\sigma$\tabularnewline
\hline 
$Z\, W\to\ell\ell\ell\nu$&
$3$&
$1000$&
$10$&
$>0.9995$ CL\tabularnewline
\hline 
$m_{h}=120$:  $W\, h\to\ell\nu b\bar{b}$ &
$3$&
$300$&
$2.4$&
$>0.9995$ CL\tabularnewline
\hline 
$m_{h}=150$: $W\, h\to(jj)\,\ell\nu\,(jj)$ &
$3$&
$300$&
$4$&
$\gg0.9995$ CL\tabularnewline
\hline
\end{tabular}
\end{center}
\end{table}

It is instructive to 
compare our analysis to the previous ones, starting with various spin-$1$
states, in order to 
illustrate 
the complementarity of the various studies.
The KK gluon has the largest cross-section in this framework, but 
it decays mostly into 
$t \bar{t}$ which results in exclusively jetty final states, 
even if the $W$ from the top decays to leptons
(due to the high degree of collimation of decay products of the top quarks).
On the other hand, KK $W$ decays into $WZ$ can result in clean, purely leptonic and fully
reconstructible final states, albeit with a small BR
which in the end does not result in the reach being larger in this
channel.\footnote{In more detail,
the ability to reconstruct the $WZ$ invariant mass makes the
$S/B$ larger, but the effect is diluted a bit when
we consider $S/\sqrt{B}$.
The effect of a smaller BR of $WZ$ vs. $WW$ to leptons
cancels in $S/B$, but still tends to reduce $S/\sqrt{B}$.
So, the net result is a significantly larger
$S/B$ for the former case, but $S/\sqrt{B}$ is not larger
by as much. Another issue is that the KK $W$ can decay into one KK and
one zero-mode fermion in the set-ups that we considered (with
decays to {\em two} light KK fermions being kinematically forbidden in
the cases that we study). The presence of this channel dilutes the
BR to $WZ$ for the KK $W$.
Such a decay channel is suppressed for the KK $Z$ since the
light KK fermion comes from a {\em different} $5D$ fermion field than the
zero-mode fermion and thus  
$U(1)$ gauge bosons (including $U(1)$ subgroups of
non-abelian gauge multiplets)
such as the $Z$ cannot couple these two fermions.}
In contrast, 
KK $Z$ decays into $WW$ can also result in purely leptonic
final states, but the invariant mass is not reconstructible in this case.
The
semileptonic analogues of these decays for KK $W/Z$ (i.e.,
one $W/Z$ decaying leptonically and the
other hadronically) are on a similar footing to KK gluon
in terms of cleanness since the
detection of the highly collimated hadronically decaying $W/Z$
requires discriminating it from the QCD jet background
(just like for highly collimated top quarks from decays of the KK gluon):
in our analysis
a jet mass cut was used for this purpose.
Finally, KK $Z$ decays to top pairs are swamped
by KK gluon background\footnote{assuming a small mass splitting between 
the KK $Z$ and the KK gluon as 
in the simplest models with no brane-localized kinetic terms for bulk
gauge fields}, but KK $W$ decays to $t \bar{b}$ do not have this problem if
the background from KK gluon to $t \bar{t}$ with one
highly boosted top faking a bottom can be reduced, for example again
by using jet mass as we did here.

We reiterate that further development of techniques for
detecting highly boosted $W/Z$ jets and similarly
vetoing a highly boosted top faking a bottom can improve the reach
for charged (and also neutral) EW states.
Another feature 
we would like to mention is that 
there are two extreme possibilities for the profiles
of top/bottom quarks which are relevant
for the KK $W$ search, 
namely, where the RH or LH top is
localized near the so called TeV brane in the extra dimension
(while the other chirality has close-to-flat profile) -- the point is
that 
either $t_R$ or $t_L$ must be localized near the TeV
brane in order to obtain the large top mass.
The first possibility is 
favored by flavor
precision tests, whereas EW precision tests have a milder preference
for the second. 
Note 
that the
KK modes are also localized near the TeV brane.
Hence, the coupling of the KK $W$ (and hence the
BR) to LH $t$ (and $\bar{b}$)\footnote{The decays of KK $W$ to $t_R$ and
$b_R$ are usually suppressed since $b_R$ is localized near 
the Planck brane.} is suppressed or enhanced
in the two cases and thus vice versa for BR of the other channels with 
significant couplings to the KK $W$, i.e., $WZ$ and $Wh$, making the
two search channels (i.e., $t \bar{b}$
and $WZ/h$) complementary in the case of the KK $W$. These two choices
for top profiles make
less of a difference for the KK $Z$ search
since the KK $Z$ always
has substantial BR to decay into SM top pairs (of
whichever chirality -- LH or RH -- is localized near the TeV brane).

For a complete perspective
of this research program, we now comment on the other spin states.
The spin-$2$ KK graviton is typically heavier than the
spin-1 states and thus has an even smaller production cross-section.
Its 
decays to $t \bar{t}$ are {\em not} likely to be swamped
by those of the KK gluon due to the different mass, but one faces 
the (even more difficult) challenge of identifying highly 
boosted top quarks.
Decays to $ZZ$ followed by leptons are possibly the cleanest and
can moreover be fully reconstructed, but suffer from a very small BR.
In contrast, decays to 
$WW$ cannot be reconstructed in the fully leptonic case
(just like in the case of KK $Z$) and challenges for the semileptonic 
case are similar to KK $W/Z$ from QCD background.

As far as KK fermions are concerned, the masses of 
the KK excitations of top/bottom (and their other gauge-group partners) in some models
(where the $5D$ gauge symmetry is extended beyond that in the SM)
can be (much) smaller than gauge KK modes, enhancing the prospect
for their discovery. 
In fact, the other heavier (spin-1 or 2) KK modes can decay
into these light KK fermions, resulting in perhaps more distinctive
final states for the heavy KK's than the pairs of $W/Z$ or top quarks
that have been studied so far -- for a recent 
such study for KK gluon, see Ref.~\cite{Carena:2007tn}.
A few studies of signals for the heavier KK fermions 
and of the radion have also been done.
We leave more detailed studies of heavier KK fermions and radion
(as
well as the
rather model-dependent decays of heavier KK
$Z/W$/graviton into lighter KK's or radion) for future work.
 
We would like to emphasize
that the signals we studied in
this paper (and the previous one
on neutral gauge bosons) might
actually be valid for a wider class of non-supersymmetric models
of EWSB. For example,
based on AdS/CFT correspondence
discussed in Sec.~\ref{TC},
it is clear that any kind of $4D$ strong dynamics involved in EWSB
will (in general) share many of the features
of the $5D$ model.
Also, the parameter
space of 
little Higgs models which satisfies EWPT 
corresponds to the $W^{ \prime }$, $Z^{ \prime }$ being
weakly coupled to light fermions and strongly coupled to
Higgs (including longitudinal $W/Z$) \cite{little},
just like in the $5D$ models we studied here. Moreover,
some UV completions of little Higgs involve $4D$ strong dynamics which
might have a dual warped extra dimensional description. Thus,
little Higgs models and EWSB models with strong dynamics
are likely to have LHC signals similar to the
ones we have studied.
Note however that the flavor structure of the
warped extra dimension is different than in traditional technicolor
models so that the decays of KK $W/Z$ to top/bottom are new 
features. 
We would also like 
to point out that the jet mass cut for semileptonic
decays of $WZ$ or $WW$ from decay of heavy $W/Z$ has not been
studied in detail in these other contexts (technicolor or little Higgs).

In more generality, the point is that there is a class of
non-supersymmetric extensions
of the SM without a symmetry (analogous to
$R$-parity in SUSY) which allow tree-level exchange
of new particles to contribute to (purely) SM operators,
resulting in strong constraints from 
precision tests, typically a few TeV
mass for the new particles. Moreover, in many such
models, the top/bottom quark and Higgs,
including the longitudinal $W/Z$, couple strongly
to the new particles since all these particles
are closely associated with EWSB. On the other
hand, the coupling of the new states to light fermions
is typically weak, in part 
based on considerations of flavor and
EW precision tests. Thus, a large class of non-supersymmetric
models faces challenges similar to the warped extra dimension
framework that we studied here, namely, production of
the new states tends to be 
suppressed and decays are mostly to top quarks/$W$/$Z$,
that too highly boosted. In summary, the techniques we developed
in this paper might be useful for obtaining signals for a wider class of 
models, beyond warped extra dimensions.

\section*{Acknowledgments}
We would like to thank H.~Davoudiasl, 
D.~E.~Kaplan, W.~Kilgore, F.~Paige, G.~Perez, Z.~Si, C.~Sturm, M.~Strassler and
R.~Sundrum for discussions, and, A.~Belyaev for help with CalcHEP, and 
S.~Mrenna and P.~Skands for help with Pythia.
KA is supported in part by NSF grant No. PHY-0652363.
%
SG and AS are supported in part by the DOE grant
DE-AC02-98CH10886 (BNL).
TH and G.-Y.H are supported in part by a DOE grant
DE-FG02-95ER40896 and in part by the Wisconsin Alumni Research Foundation,
and G.-Y.H is also supported by DOE grant DE-FG02-91ER40674 and
by the U.C. Davis HEFTI program. 
%

\appendix

\section{Couplings and Mixing Angles}
\label{Coupl.APP}

Here we collect from Ref.~\cite{Agashe:2007ki}, expressions for couplings 
and mixing angles. We focus mainly on the fermion representation 
with the custodial symmetry protecting $Zb\bar b$.
For our numerical study, we assume $g_L = g_R$ throughout.
The mixing angles and couplings are related through (with $s\equiv \sin()$ and $c\equiv cos()$)
\bea
g^\prime &=& \frac{g_X g_R}{\sqrt{g_R^2 + g_X^2}} \ \ , \ \ s^\prime = \frac{g_X}{\sqrt{g_R^2 + g_X^2}} \ \ , \ \ c^\prime = \sqrt{1-{s^\prime}^2} \ , \\
e &=& \frac{g_L g^\prime}{\sqrt{g^{\prime 2} + g_L^2}} \ \ , \ \ s_W = \frac{g^\prime}{\sqrt{g^{\prime 2} + g_L^2}} \ \ , \ \ c_W = \sqrt{1-s_W^2} \ , \\
g_Z &=& g_L/c_W \ \ , \ \ g_{Z^\prime} = g_R/c^\prime \ .
\eea
For the case $g_R = g_L$, we have $s^\prime = 0.55$, $c^\prime = 0.84$.

The various mixing angles are as explained in Ref.~\cite{Agashe:2007ki} and 
we repeat below a few relevant ones.
As explained in Apps.~A and B of Ref.~\cite{Agashe:2007ki}, EWSB induces a mixing between 
$Z^{(0)} \leftrightarrow \zp$ (with mixing angle $\theta_{01}$) 
and $Z^{(0)} \leftrightarrow \zx$ (with mixing angle $\theta_{01X}$). 
To leading order in $\mz/\mzpri$ these mixing angles are given by
\bea
\sin{\theta_{01}} \approx \left(\frac{\mz}{\mzp} \right)^2 \sqrt{k \pi r_c} \ ,\label{sth01.EQ} \\
\sin{\theta_{01X}} \approx -\left(\frac{\mz}{\mzx} \right)^2 \left(\frac{g_{Z^\prime}}{g_Z}\right) c^{\prime\,2} \sqrt{k \pi r_c} \ .
\label{sth01X.EQ}
\eea
For example, for $\mzpri = 2$~TeV, $s_{01} = 0.013$ and $s_{01X} = -0.01$.

EWSB similarly induces mixing in the charged $W^\pm$ sector i.e. mixing between 
$W \leftrightarrow \wpri$, with mixing angle given by
\bea
\sin{\theta_{0L}} \approx \left(\frac{\mw}{\mwpL} \right)^2 \sqrt{k \pi r_c} \ , \label{sth0L.EQ} \\
\sin{\theta_{0R}} \approx -\left(\frac{\mw}{\mwpR} \right)^2 \left(\frac{g_R}{g_L}\right) \sqrt{k \pi r_c} \ . \label{sth0R.EQ}
\eea
For example, for $\mzpri = 2$~TeV, $s_{0L} \approx 0.01$ and $s_{0R} \approx - 0.01 $.

EWSB also induces $\zp\leftrightarrow \zx$ mixing, with mixing angle given by
\beq
\tan{2\theta_1} = \frac{-2 \mz^2 (g_{Z^\prime}/g_Z) c^{\prime 2} k\pi r_c }{(\mzx^2 - \mzp^2) + 
\mz^2\left( (g_{Z^\prime}/g_Z)^2 c^{\prime 4} - 1\right) k\pi r_c} \ .
\eeq
For example, for $\mzp = 2000; \mzx = 1962$~GeV, this implies that $s_1 = 0.48$, $c_1 = 0.88$.
After this mixing, we will refer to the mass eigenstates as $\zpt$ and $\zxt$.

EWSB similarly induces $\wpL \leftrightarrow \wpR$ with  mixing angle given by
\beq
\tan{2\theta_1^c} = \frac{-2 \mw^2 (g_R/g_L) k\pi r_c }{(\mwpR^2 - \mwpL^2) + 
\mw^2\left( (g_R/g_L)^2 - 1\right) k\pi r_c} \ .
\eeq
For example, for $\mwpL = 2000; \mwpR = 1962$~GeV, this implies that $s^c_1 = 0.6$, $c^c_1 = 0.8$.
After this mixing, we will refer to the mass eigenstates as $\wpLt$ and $\wpRt$,
and for notational easy we will just denote them as $\wpLnt$ and $\wpRnt$ 
respectively.

\section{Couplings of $W^{ \prime }$}
\label{coupling}

The electroweak gauge group in the bulk is $SU(2)_L \times SU(2)_R \times
U(1)_X$,
with hypercharge being a
linear combination of
$U(1)_R$ and $U(1)_X$.
The extra $SU(2)_R$ (relative to the SM) ensures
suppression of contribution to the
EWPT (specifically an
observable called the $T$ parameter). Hence, we obtain
$2$ charged KK towers (before EWSB) --
one from each $SU(2)$ group in the bulk.
We will restrict to the 1st KK modes only
in each tower
and denote them by
$W_{ L1 }$ and $W_{ R1 }$, respectively.
EWSB mixes the
2 and the resulting mass eigenstates
are denoted by $\tilde{W}_{ R1 }$ and $\tilde{W}_{ L1 }$.

We work in the approximation $ (k\pi r_c) \, m_W^2/M_{KK}^2 \ll 1 $.  

\subsection{$\wpri$ coupling to fermions}
We show below the fermion representations under $SU(2)_L \otimes SU(2)_R \otimes U(1)_X$, 
denoted as $(L,R)_X$. We take the left handed quarks of the first and second generation, and the 
left-handed leptons to be doublets under $SU(2)_L$. This specifies the interaction of these fields
with $\wpL$.
The $\wpR$ couplings to first and second generation quarks, 
right-handed bottom quark and leptons are negligibly small since the $\wpR$ profile is 
suppressed near the Planck brane where these fermion fields are peaked. 
To have the custodial symmetry protection of the $ Zb\bar b $ coupling~\cite{custodial2}, 
we take the third generation left-handed quarks to be in the representation
\beq
Q_L^3 = \bmat q_L^3 & {q_L^\prime}^3 \emat  =  \bmat t_L & \chi_L \\ b_L & T_L \emat \to (2,2)_{2/3} \ ,
\label{Q3Ldef.EQ}
\eeq
where $ \chi_L , T_L $ are taken to have $ (-+) $ boundary conditions (BC) with no zero-modes\footnote{All
SM fermions have (++) BC since they are zero modes.}.
We have $ Q(\chi_L) = 5/3 $ and $ Q(T_L) = 2/3 $.
To accommodate the large top and bottom mass difference we take it that $ t_R $ and $ b_R $ do not belong
to the same $SU(2)_R$ multiplet.
We consider two cases for the $ t_R $ representations
\bea
{\rm Case (i)} &:&\  t_R \to (1,1)_{2/3} \ ,  \nonumber \\
{\rm Case (ii)} &:&\ t_R \to (1,3)_{2/3} \oplus (3,1)_{2/3} = \bmat \chi_R^{\prime\prime} \\ t_R \\ B_R^{\prime\prime} \emat \oplus \bmat \chi_R^{\prime\prime\prime} \\ T_R^{\prime\prime\prime} \\ B_R^{\prime\prime\prime} \emat  \ ,
\label{tRdef.EQ}
\eea
where the exotic fermions have $(-+)$ BC with no zero-modes, and the fermions in the $(3,1)$ representation 
are not discussed further in our work here since the $\wpri$ decay to a pair of them is kinematically 
forbidden. 
For Case (i), $t_R\to (1,1)$, the electroweak precision tests (EWPT) are better 
satisfied~\cite{EWPTmodel}
for $c_{ Q_L^3 } = 0$ and $c_{ t_R } = 0.4$, i.e., $Q_L^3$ peaked closer to the TeV brane,
while for Case (ii), $t_R\to (1,3)$, for 
$c_{ Q_L^3 } = 0.4$ and $c_{ t_R } = 0$, i.e., $c_{t_R}$ peaked closer to the TeV brane. 
After including the charges and the overlap integrals, the largest effective coupling of third generation
fermions to gauge KK modes in Case (i) would be to $Q_L^3$, being larger than that in Case (ii),
which would be to $t_R$.   
Consequently, while on the one hand new gauge KK induced FCNC contributions would be larger in 
Case (i) and hence more problematic for the simplest constructions, 
on the other hand collider signals would be larger compared to Case (ii).

The fermion couplings to $\wpri$ depend on various mixing angles summarized in App.~\ref{Coupl.APP}.
The couplings also depend on the overlap integrals which we give next.
We note that $\wpL$ has (++) while $\wpR$ has $(-+)$ BC.
The overlap integrals of a $\wpri$ with two fermions are given in Table~\ref{ovlap_ffG.TAB}.
We represent by ${\cal I}^{xx}_{yy,zz}$ the overlap integral of the $\wpri$ having $xx$ BC with
two fermion fields one with $yy$ and the other with $zz$ BC's.  
For instance, ${\cal I}^{++}_{++,++}$ is the overlap integral of the $\wpL$ 
with two fermions both with (++) BC, and,
${\cal I}^{++}_{++,-+}$ the overlap integral of the $\wpL$ with one fermions with (++) BC 
and the other with $(-+)$ BC. 
Similarly, $ {\cal I}^{-+}_{yy,zz} $ represents the overlap integral of the $\wpR$ with two fermions.
Due to the orbifold $Z_2$ symmetry, we have $I^{++}_{++,-+} = 0$ etc., and we show only the
nonzero ones in the table.
\begin{table}[h]
\begin{center}
\caption{Values of $\psi \psi \wpri$ overlap integrals for:
Case (i), $t_R \to (1,1)$, $c_{ Q_L^3 } = 0$ and $c_{ t_R } = 0.4$ (upper table), 
and, 
Case (ii), $t_R \to (1,3)$, $c_{ Q_L^3 } = 0.4$ and $c_{ t_R } = 0$ (lower table). 
All the other $c$'s $> 0.5$.
We take $\xi = \sqrt{k\pi r_c} = 5.83$.
All SM fermions have (++) BC, "exotic" BSM fermions have $(-+)$, $\wpL$ has (++), and,
$\wpR$ has $(-+)$ BC. 
 \label{ovlap_ffG.TAB}}
\begin{tabular}{|c||c|c|c|}
\hline 
$ c_{ Q_L^3 } = 0 , c_{ t_R } = 0.4 $ & $Q_{L}^{3}$ & $t_{R}$ & other fermions
\tabularnewline
\hline
\hline 
${\cal I}^{++}_{++,++}$   & $- \frac{ 1.13 }{ \xi } + 0.7 \xi \approx 3.9$ & $- \frac{ 1.13 }{ \xi } + 0.2 \xi \approx 1$   & $- \frac{ 1.13 }{ \xi } \approx -0.2$
\tabularnewline
\hline 
${\cal I}^{++}_{-+,-+}$ & $\xi $ & $\xi$ & $ - $ 
\tabularnewline
\hline 
${\cal I}^{-+}_{++,-+}$ & $ 0.8 \xi \approx 4.6 $ & $ 0.4 \xi \approx 2.3 $  & $\approx 0$
\tabularnewline
\hline
\end{tabular}
\begin{tabular}{|c||c|c|c|}
\hline 
$ c_{ Q_L^3 } = 0.4 , c_{ t_R } = 0 $ & $Q_{L}^{3}$ & $t_{R}$ & other fermions
\tabularnewline
\hline
\hline 
${\cal I}^{++}_{++,++}$ & $- \frac{ 1.13 }{ \xi } + 0.2 \xi \approx 1$ & $- \frac{ 1.13 }{ \xi } + 0.7 \xi \approx 3.9$  & $- \frac{ 1.13 }{ \xi } \approx -0.2$
\tabularnewline
\hline 
${\cal I}^{++}_{-+,-+}$ & $\xi$ & $\xi $& $ - $ 
\tabularnewline
\hline 
${\cal I}^{-+}_{++,-+}$ & $ 0.4 \xi \approx 2.3 $ & $ 0.8 \xi \approx 4.6 $ & $\approx 0$
\tabularnewline
\hline
\end{tabular}
\end{center}
\end{table}

We note that the mass of the $(-+)$ fermion is lighter than $M_{\wpLt}$ for $c < 1/2$. In particular for $c=0.4$ it is about $0.9 M_{\wpLt}$, and for $ c=0$ it is about $0.6 M_{\wpLt}$.
The first KK excitation of the (++) fermions are typically heavier than $M_{\wpLt}$ (being equal at $c=1/2$).

The $\wpri$ coupling (Feynman rule) to fermions is given by 
\beq
\overline{u_L} d_L \{\wpLt^+ , \wpRt^+ \} \ : \ i \frac{g_L}{\sqrt{2}} \{c^c_1, s^c_1 \} {\cal I}^{++}_{u_L d_L} \ , 
\label{udWp.EQ}
\eeq
where $u_L$ ($d_L$) denotes first and second generation up- (down-) type fermions. 
The third generation left-handed fermion couplings are give by
\bea
 \overline{t_L} b_L \{\wpLt^+ , \wpRt^+ \} \ &:& \ i \frac{g_L}{\sqrt{2}} \{c^c_1, s^c_1 \} {\cal I}^{++}_{t_L b_L} \ ,  \nonumber \\
 \overline{\chi_L} T_L \{\wpLt^+ , \wpRt^+ \} \ &:& \ i \frac{g_L}{\sqrt{2}} \{c^c_1, s^c_1 \} {\cal I}^{++}_{\chi_L T_L} \ ,  \nonumber \\
 \overline{\chi_L} t_L \{\wpLt^+ , \wpRt^+ \} \ &:& \ i \frac{g_R}{\sqrt{2}} \{s^c_1, -c^c_1 \} {\cal I}^{-+}_{\chi_L t_L} \ ,  \nonumber \\
\overline{T_L} b_L \{\wpLt^+ , \wpRt^+ \} \ &:& \ i \frac{g_R}{\sqrt{2}} \{s^c_1, -c^c_1 \} {\cal I}^{-+}_{T_L b_L} \ .
\eea

For Case (i), $t_R \to (1,1)$, it does not interact with the $\wpri$ as already mentioned. 
For Case (ii), $t_R \to (1,3)$ its interaction with the $\wpri$ is given as
\bea
\overline{\chi_R^{\prime\prime}} t_R \{\wpLt^+, \wpRt^+\} \ &:& \ -i \frac{g_R}{2} \{-s^c_1, c^c_1  \}  {\cal I}^{-+}_{\chi_R^{\prime\prime} t_R} \ , \nonumber \\
\overline{t_R} B_R^{\prime\prime} \{\wpLt^+, \wpRt^+\} \ &:& \ i \frac{g_R}{2} \{-s^c_1, c^c_1  \}  {\cal I}^{-+}_{t_R B_R^{\prime\prime}} \ .
\eea

\subsection{$\wpri$ coupling to two SM gauge bosons}
In order to derive the triple gauge boson coupling we start with the
KK basis Lagrangian terms (keeping in mind ${W_R^\pm}^{(0)}\equiv 0$)
\beq
{\cal L} \supset - g_L {W^3_L}^{(0)} {W^+_L}^{(0)} {W^-_L}^{(0)}     
                 - g_L {W^3_L}^{(1)} {W^+_L}^{(1)} {W^-_L}^{(0)}   
                 - g_L {W^3_L}^{(0)} {W^+_L}^{(1)} {W^-_L}^{(1)}   
                 - g_R {W^3_R}^{(0)} {W^+_R}^{(1)} {W^-_R}^{(1)}  \ .
\eeq  
Writing this in the mass eigenbasis results in the triple gauge boson couplings (Feynman rules). 
The $ A \, {\wpri}^\pm W^\mp $ coupling is zero. 
The $Z$ couplings (Feynman rules) are given by 
\bea
Z \, \wpLt^+ W^-  \ &:& \  - i g_L c_W \left[ -s_{0R}\, s^c_1 \left(\frac{g_R}{g_L}\frac{s_W}{c_W} s^\prime  + 1 \right) - s_{01} c^c_1 \right] \ , \nonumber \\
Z \, \wpRt^+ W^-  \ &:& \  - i g_L c_W \left[ s_{0R}\, c^c_1 \left(\frac{g_R}{g_L}\frac{s_W}{c_W} s^\prime  + 1 \right) - s_{01} s^c_1 \right] \ , 
\eea
and for comparison we note that the SM triple gauge boson coupling is given as
\\ \hspace*{2cm} $\{A,Z\}\, W^+ W^-  \ : \  - i g_L \{s_W,c_W\}$.

\subsection{$\wpri$ coupling to $W$ and Higgs}
Starting from Eq.~(44) of Ref.~\cite{Agashe:2007ki} we obtain the couplings to the Higgs 
by making the substitution $v \to (v+h)$ which results in the couplings (Feynman rules)
\bea
\{ \wpLt^+ , \wpRt^+  \} W^- h \ &:& i \ \frac{2 m_W^2}{v} \sqrt{k\pi r_c} 
\left\{ \left( c^c_1 + \frac{g_R}{g_L} s^c_1 \right) ,  \left( s^c_1 - \frac{g_R}{g_L} c^c_1\right)  \right\} \ ,  \nonumber \\
\{ \wpLt^+ , \wpRt^+  \} W^- h h \ &:& i \ \frac{2 m_W^2}{v^2} \sqrt{k\pi r_c} 
\left\{ \left( c^c_1 + \frac{g_R}{g_L} s^c_1 \right) ,  \left( s^c_1 - \frac{g_R}{g_L} c^c_1\right)  \right\} \ ,
\eea
where the $hh$ couplings include a symmetry factor of 2.


\end{document}